\documentclass[11pt]{article}
\usepackage{geometry}

\usepackage{amsmath,amssymb,amsthm,booktabs}
\usepackage{mathrsfs}
\usepackage[OT1]{fontenc}
\usepackage[utf8]{inputenc}
\usepackage[colorlinks,citecolor=blue,urlcolor=blue]{hyperref}
\usepackage{txfonts}
\usepackage{indentfirst}
\usepackage{multirow}
\usepackage{float,subfig}

\usepackage{bm}
\usepackage{euscript}
\usepackage{graphicx}
\usepackage{multicol}
\usepackage[usenames,dvipsnames,svgnames,table]{xcolor}

\usepackage[round]{natbib}
\numberwithin{equation}{section} \theoremstyle{plain}
\newtheorem{theorem}{Theorem}[section]
\newtheorem{lemma}{Lemma}[section]

\newtheorem{proposition}{Proposition}[section]

\newtheorem{example}{Example}

\newtheorem{remark}{Remark}[section]

\numberwithin{equation}{section}

\begin{document}

\title{Discrete time multi-period mean-variance model: Bellman type strategy and
Empirical analysis
\footnotemark[1]}
\author{Shuzhen Yang
\footnotemark[2]
\footnotemark[3]}

\renewcommand{\thefootnote}{\fnsymbol{footnote}}

\footnotetext[1]{

\textbf{\ \ Journal of Economic Literature classification Numbers}: C61; D81; G11.

\ \ \textbf{MSC2010 subject classification}: 90C39; 93E20; 49L10.
}
\footnotetext[2]{Shandong University-Zhong Tai Securities Institute for Financial Studies, Shandong University, PR China, (yangsz@sdu.edu.cn).}
\footnotetext[3]{This work was supported by the National Natural Science Foundation of China (Grant No.11701330) and Young Scholars Program of Shandong University.}

\date{}
\maketitle

\renewcommand{\thefootnote}{\alph{footnote}}

 \textbf{Abstract}: In this paper, we attempt to introduce the Bellman principle for a discrete time multi-period mean-variance model. Based on this new take on the Bellman principle, we obtain a dynamic time-consistent optimal strategy and related efficient frontier.  Furthermore, we develop a varying investment period discrete time multi-period mean-variance model and obtain a related dynamic optimal strategy and an optimal investment period. This paper compares the highlighted dynamic optimal strategies of this study with the $1/n$ equality strategy, and shows that we can secure a higher return with a smaller risk based on the dynamic optimal strategies.

\addcontentsline{toc}{section}{\hspace*{1.8em}Abstract}

\section{Introduction}
Since the foundational work of \cite{Ma52,Ma59} on the portfolio selection problem, the mean-variance investment model has been used to balance the return and risk of the wealth of portfolios (Also see \cite{Ma14}). The single-period mean-variance model, which is used presently, was developed to include an explicit solution of the optimal strategy and its application in real markets. (Further see \cite{Me72}). Apart from the single-period framework, many authors consider studying the multi-period mean-variance model, which optimizes the multi-period objectives with a dynamic strategy. A lot of literature discusses the multi-period mean-variance model, including both the discrete and continuous time cases. For a given investment period $T$, the investor wants to minimize variance in the portfolio's  wealth at $T$ with conditions regarding the return on investment. Because the variance does not satisfy the iterated-expectation property, it is difficult to find a dynamic, time-consistent, optimal strategy for the multi-period mean-variance model. There are two kinds of optimal strategies for solving the multi-period mean-variance model: one is called the pre-committed strategy which is derived by optimizing the variance of wealth at $T$, and the second is called the game-theoretic strategy which is derived by defining a local maximum principle.

For the pre-committed strategy, a continuous time mean-variance model with one risky asset stock and bond was considered in \cite{R89} and subsequently the optimal strategy was given. When combining initial time and terminal time, \cite{BP98} developed the  portfolio strategies for the related mean-variance model. By embedding the multi-period mean-variance problem into a multi-objective optimization framework, \cite{LN00} found a related dynamic optimal strategy. Later, many authors began to study the mean-variance model in continuous time.  Based on a similar technique in \cite{LN00}, the continuous-time mean-variance problem was studied in \cite{ZL00} and  an optimal strategy and efficient frontier were given. In a straightforward manner,  a cost-efficient approach to the optimal portfolio selection for the mean-variance problem was proposed in \cite{D88}. Employing the cost-efficient approach, \cite{BS14} solved the problem of a mean-variance in the presence of a benchmark. We can see that the optimal strategy was suggested by the cost-efficient approach, which is consistent with the results of \cite{ZL00}. Furthermore, we refer the reader to \citep{BJPZ05,DXZ10,LZ02,L04,X05,BJM18} for more details of the mean-variance problem in continuous time.

Based on the optimal control problem of a stochastic differential equation of mean-field type, the stochastic maximum principle to the mean-variance portfolio selection problem was investigated in \cite{AD11}, and the related optimal strategy coexists with that in \cite{ZL00}. Furthermore, an integral form stochastic maximum principle for general mean-field optimal control systems was established in \cite{L12}, and was applied to solve a mean-field type linear quadratic stochastic control problem, also see \cite{BDL11}.  Based on the mean field approach, the continuous time mean-variance portfolio optimization problem  was studied in \cite{FL16}, which obtained a related pre-committed strategy. The optimal control of a general stochastic McKean-Vlasov equation was studied in \cite{PW17}, and the dynamic programming principle for the value function was established in the Wasserstein space of probability measures. Furthermore,  \cite{PW17} solved the linear-quadratic stochastic McKean-Vlasov control problem and an interbank systemic risk model with common noise were investigated, also see \cite{PW18}.  The explicit solution for the optimal robust portfolio strategies in the case of uncertain volatilities was given in \cite{IP19}, which coincides with those in \cite{ZL00} and \cite{FL16}.

For the game-theoretic strategy, a dynamic method is given in \cite{BC10}, which is used to study the mean-variance model by introducing an adjustment term in the objective.  A general time-inconsistent stochastic linear-quadratic control problem was established in \cite{HJZ12}, and an equilibrium, instead of optimal control was defined. (Further see \cite{Y12}). In addition to this, \cite{HJZ12} considered a pre-committed strategy for the mean-variance model. The large population stochastic dynamic games and the Nash Certainty Equivalence based control laws was investigated in \cite{HCM07}. We refer the readers to  \citep{BSY16,BSY13,BMZ14,BKM17,DJSX19} for more details of game-theoretic approach.

Recently, \cite{Y20} investigated a new method to solve the multi-period mean-variance problem in continuous time. Let $X^{\pi}(\cdot)$ denote the wealth of the investor in the investment time interval $[t,T]$ with the initial time $t$, where $\pi(\cdot)$ is the related strategy. Note that, the variance $\mathrm{Var}[X^{\pi}(T)]$ does not satisfy the iterated-expectation property, which deduces that one cannot use the dynamic programming principle to solve the mean-variance problem in multi-period mean-variance model. To solve the above problem, \cite{Y20} introduced a deterministic process $Y^{\pi}(\cdot)$ to represent the mean process $\mathbb{E}[X^{\pi}(\cdot)]$ such that the variance satisfies Bellman dynamic programming principle. Based on the idea developed in \cite{Y20}, we want to investigate the discrete time multi-period mean-variance problem and consider the related empirical analysis in this study. Note that, we cannot use the  It\^{o} formula and partial differential equation tool which were used in \cite{Y20} to solve the multi-period mean-variance problem in discrete time case. By introducing a deterministic process $Y^{\pi}(\cdot)$, which is equal to the expectation of $X^{\pi}(\cdot)$ with initial value $y\in \mathbb{R}$, the related cost functional is given as follows:
\begin{equation}\label{in-cost}
\tilde{J}(t,x,y,\mu;\pi(\cdot))=\mu \mathbb{E}[\big(X^{\pi}(T)-Y^{\pi}(T)\big)^2]-\mathbb{E}[X^{\pi}(T)],
\end{equation}
where $\mu$ is the risk aversion coefficient and $X^{\pi}(T)$ with initial value $x\in \mathbb{R}$. Hence, we can separate the process $Y^{\pi}(T)$ from the cost functional (\ref{in-cost}), by defining the value function:
$$
V^{\mu}(t,x,y)=\inf_{\pi\in\mathcal{A}_t^{T-1}}\tilde{J}(t,x,y,\mu;\pi(\cdot)),
$$
where $\mathcal{A}_t^{T-1}$ is the set of all adapted strategies. First, we can establish the Bellman principle for value function $V^{\mu}(t,x,y)$, based on which we can obtain the Bellman type dynamic time-consistent optimal strategy and dynamic efficient frontier for the value function $V^{\mu}(t,x,y)$. We denote the related optimal strategy as STRATEGY I. In particular, when the number of investment period $N \to \infty$, the Bellman type optimal strategy converges to the optimal strategy given in \cite{Y20}.

To reduce the variance of the classical mean-variance model in continuous time, \cite{Y19} proposed a varying investment period mean-variance model with a constraint on the mean value of the portfolio's wealth, which moves  with the varying investment period. In this study, we introduce a varying investment period mean-variance model in discrete time. For a given deterministic investment period $\tau\in \mathbb{N}^+$, we consider a varying investment period structure for the classical discrete mean-variance model:
\begin{equation*}
\tau^{\pi}=\inf\{s :\ \mathbb{E}[X^{\pi}(s)]\geq g(\tau),\ 0<s\leq \tau \}\bigwedge \tau,
\end{equation*}
where $a_1\bigwedge a_2,\ a_1,a_2\in \mathbb{R}$ means $a_1\bigwedge a_2=\min(a_1,a_2)$, and $g(\cdot)$ is the varying expected return on investment $X^{\pi}(\cdot)$. Based on this varying investment period $\tau^{\pi}$, we obtain the related Bellman type time-consistent optimal strategy and optimal investment period. We denote the related optimal strategy as STRATEGY II.

Based on the out-of-sample performance of the sample-based mean-variance model, \cite{DGU09} suggests that the $1/n$ equality strategy should serve as a first obvious benchmark. To compare the efficient of STRATEGY I, STRATEGY II and $1/n$ equality strategy, we use the daily data of NASDAQ and Dow Jones from the period of Aug. 03, 2009 to Aug. 02, 2019 to construct the portfolio investments of STRATEGY I, STRATEGY II, and the $1/n$ equality strategy. Here, we show that we can obtain a yearly return of $270.78\%$ with Sharpe ratio 0.8077 based on STRATEGY I, a yearly return of $249.12\%$ with Sharpe ratio 0.6287 based on STRATEGY II, and a yearly return of $11.65\%$ with Sharpe ratio 0.7370 based on the $1/n$ equality strategy.

The remainder of this paper is organized as follows: in Section \ref{mvm-1}, we formulate the discrete time multi-period mean-variance model. In Section \ref{vmvm-1}, based on the dynamic programming principle of the value function, we establish a Bellman type dynamic time-consistent optimal strategy and a dynamic time-consistent relationship between the mean and variance. Furthermore, a varying investment period discrete mean-variance model is investigated in Section \ref{vmvm-1}. To compare the efficiency of the Bellman type dynamic optimal strategy, as deduced in Section \ref{vmvm-1}, with the $1/n$ equality strategy, we construct the portfolio investment for the index NASDAQ and Dow Jones by the strategies given in Section \ref{ea}. Finally, we conclude  the paper in Section \ref{con}.

\section{Multi-period mean-variance model}\label{mvm-1}
\subsection{Discrete mean-variance model}
Given a complete filtered probability space $(\Omega,\mathcal{F},P;\{ \mathcal{F}(s)\}_{s\geq
t})$, and $W(\cdot)$, which is a $d$-dimensional standard Brownian motion defined on which with $W(t)=0$,  where $ \mathcal{F}(s)$ is the $P$-augmentation of the
natural filtration generated by $(W(t),W(t+1),\cdots,W(s)),\ t\leq s\leq T$, where $T$ is the given investment period. We consider that one risk-free bond asset and $n$ risky  stock assets are traded in the market, where the bond satisfies:
\begin{eqnarray*}
\left\{\begin{array}{rl}
\displaystyle P_0(s)& \!\!\!= P_0(s-1)r(s-1),  \\
 P_0(t) & \!\!\!= p_0,\ \ t<s\leq T,
\end{array}\right.
\end{eqnarray*}
and the $i$'th ($1\leq i\leq n $) stock asset is described by
\begin{eqnarray*}
\left\{\begin{array}{rl} \displaystyle  P_i(s) & \!\!\!=
 \displaystyle P_i(s-1)\bigg{[}b_i(s-1)+\sum_{j=1}^d\sigma_{ij}(s-1)\Delta W_j(s-1)\bigg{]},\;\;
\\
 P_i(t) & \!\!\!= p_i,\ \ t<s\leq T,
\end{array}\right.
\end{eqnarray*}
where $\Delta W_j(s-1)=W_j(s)-W_j(s-1)$, $r(\cdot)\in \mathbb{R}$ is the risk-free return of the bond, $b(\cdot)=(b_1(\cdot),\cdots,b_n(\cdot))\in{\mathbb{R}^n}$ is the expected return of the risky assets. Given initial capital $x>0$, $\displaystyle \gamma(\cdot)=(\gamma_1(\cdot),\cdots,\gamma_n(\cdot))\in \mathbb{R}^n$, where $\gamma_i(\cdot)=b_i(\cdot)-r(\cdot),\ 1 \leq i \leq n$. The investor's wealth $X^{\pi}(\cdot)$ satisfies
\begin{equation}\label{asset-1}
\left\{\begin{array}{rl}
X^{\pi}(s)  & \!\!\!=r(s-1)X^{\pi}(s-1)  +\gamma(s-1)\pi(s-1)^{\top}+\pi(s-1)\sigma(s-1) \Delta W(s-1),  \\
\!X^{\pi}(t) & \!\!\!=x,\ \ t<s\leq T,
\end{array}\right.
\end{equation}
where $\sigma(\cdot)=(\sigma_1(\cdot),\cdots,\sigma_d(\cdot))\in \mathbb{R}^{n\times d}$, $\sigma_{i}(\cdot)=(\sigma_{i1}(\cdot),\cdots,\sigma_{in}(\cdot))^{\top}$, $\pi(\cdot)=(\pi^1(\cdot),\cdots,\pi^n(\cdot))\in \mathbb{R}^{n}$ is the capital invested in the risky asset $S(\cdot)=(S_1(\cdot),\cdots,S_n(\cdot))\in \mathbb{R}^n$ and  $\pi^0(\cdot)$ is the capital invested in the bond. Thus, we have
$\displaystyle X^{\pi}(\cdot)=\sum_{i=0}^n\pi^i(\cdot)$.

In this study, we consider the following  mean-variance model:
\begin{equation}
J(t,x;\pi(\cdot))=\mathrm{Var}[X^{\pi}(T)]=
\mathbb{E}[\big{(}X^{\pi}(T)-\mathbb{E}[X^{\pi}(T)]\big{)}^2],\label{cost-1}%
\end{equation}
with the following constraint on the mean,
\begin{equation}
 \mathbb{E}[X^{\pi}(T)]=L. \label{mean-1}
\end{equation}
The set of admissible strategies $\pi(\cdot)$ is defined as:
$$
\mathcal{A}^{T-1}_t=\bigg{\{}\pi(\cdot):  \pi(s)\in L^2[\mathcal{F}_s,\ \mathbb{R}^n],\
t\leq s<T\bigg{\}}.
$$
The following assumptions are used  to obtain the optimal strategy for the proposed  model (\ref{cost-1}):

{$\textbf{H}_1$}: $r(\cdot), b(\cdot)$ and $\sigma(\cdot)$ are  deterministic functions.

{$\textbf{H}_2$}: $r(\cdot)>0$, $\gamma(\cdot)\neq 0$,  $\sigma(\cdot)\sigma(\cdot)^{\top}>\delta  \textbf{I}$, where $\delta>0$ is a given constant and $\textbf{I}$ is the identity matrix of $\mathbb{S}^{n}$, and $\mathbb{S}^{n}$ is the set of symmetric matrices.

\subsection{Bellman principle}
To obtain the Bellman type optimal strategy for the discrete time mean-variance model (\ref{asset-1}), we set the term $\mathbb{E}[X^{\pi}(\cdot)]$ as a deterministic process $Y^{\pi}(\cdot)$, which can separate the term $\mathbb{E}[X^{\pi}(\cdot)]$  from the variance  $\mathrm{Var}[X^{\pi}(T)]$. In the following, we consider the cost functional:
\begin{equation}
J(t,x,\mu;\pi(\cdot))=
\mu \mathrm{Var}[X^{\pi}(T)]-\mathbb{E}[X^{\pi}(T)],\label{cost-2}%
\end{equation}
where $\mu>0$ is the risk aversion coefficient and can be determined by the mean constraint $L$ in (\ref{mean-1}). Notice that,
$$
\mathrm{Var}[X^{\pi}(T)]=\mathbb{E}[\big{(}X^{\pi}(T)-\mathbb{E}[X^{\pi}(T)]\big{)}^2].
$$
We cannot obtain the Bellman principle for the term $[\mathbb{E}X^{\pi}(T)]^2$ because $[\mathbb{E}(\cdot)]^2$ is a nonlinear function of $\mathbb{E}(\cdot)$. To separate the expectation term $\mathbb{E}[X^{\pi}(T)]$ from the variance $\mathrm{Var}[X^{\pi}(T)]$, we introduce the following auxiliary process $Y^{\pi}(\cdot)$, where $Y^{\pi}(\cdot)$ satisfies
 \begin{equation}\label{asset-2}
\left\{\begin{array}{rl}
Y^{\pi}(s)  & \!\!\!=r(s-1)Y^{\pi}(s-1)  +\gamma(s-1)\mathbb{E}[\pi(s-1)^{\top}],  \\
\!Y^{\pi}(t) & \!\!\!=y,\ \ t<s\leq T,
\end{array}\right.
\end{equation}
Comparing  equations (\ref{asset-1}) and (\ref{asset-2}), it follows that
$Y^{\pi}(s)=\mathbb{E}[X^{\pi}(s)]$ when $x=y$, $t\leq s\leq T$.

Now, we consider the following general cost functional (\ref{cost-2}):
\begin{equation*}
\begin{array}{rl}
\tilde{J}(t,x,y,\mu;\pi(\cdot))=\mu \mathbb{E}[\big(X^{\pi}(T)-Y^{\pi}(T)\big)^2]-\mathbb{E}[X^{\pi}(T)].
\end{array}
\end{equation*}
The value function is defined as
\begin{equation}
V^{\mu}(t,x,y)=\inf_{\pi(\cdot)\in \mathcal{A}^{T-1}_t}\tilde{J}(t,x,y,\mu;\pi(\cdot)).\label{value-1}%
\end{equation}

We have the following Bellman principle for the value function $V^{\mu}(t,x,y)$. The proofs of Theorem \ref{the-1} and Theorem \ref{the-2} are given in Appendix \ref{sec:proof}.
\begin{theorem}\label{the-1}
Let Assumptions {$\textbf{H}_1$} and {$\textbf{H}_2$} hold. For any given $0\leq t\leq s< T,\ x,y\in \mathbb{R}$, we have,
\begin{equation}\label{belm-1}
V^{\mu}(t,x,y)=\inf_{\pi(\cdot)\in \mathcal{A}_t^{s-1}}\mathbb{E}[V^{\mu}(s,X^{\pi}(s),Y^{\pi}(s))].
\end{equation}
\end{theorem}

\begin{theorem}\label{the-2}
Let Assumptions {$\textbf{H}_1$} and {$\textbf{H}_2$} hold. For any given $0\leq t< T,\ x\neq y\in \mathbb{R}$,
\begin{equation}\label{pde-10}
V^{\mu}(t,x,y)=\mu (x-y)^2\bigg(\prod_{s=t}^{T-1}r(s)\bigg)^2-x\prod_{s=t}^{T-1}r(s)
-\frac{1}{4\mu}\sum_{s=t}^{T-1}\beta(s),
\end{equation}
is the solution of equation (\ref{belm-1}),
where $\beta(t)=\gamma(t)[\sigma(t)\sigma(t)^{\top}]^{-1}\gamma(t)^{\top}$, and the related optimal strategy is
$$
\pi^*(t,x,y)=\frac{1}{2\mu}\gamma(t)\big[\sigma(t)\sigma(t)^{\top}\big]^{-1}
\bigg(\prod_{s=t}^{T-1}r(s)\bigg)^{-1},\ 0\leq t<T.
$$
\end{theorem}
The notation $\prod_{s=t}^{T-1}r(s)$ means $r(t) r(t+1)\cdots r(T-1)$. If $T-1<t$, we set $\prod_{s=t}^{T-1}r(s)=1$.

\begin{remark}\label{re-2-1}
For any given initial time and states $(t,x,y)$, from Theorem \ref{the-2}, we can obtain the optimal strategy
$$
\pi^*(t,x,y)=\frac{1}{2\mu}\gamma(t)\big[\sigma(t)\sigma(t)^{\top}\big]^{-1}
\bigg(\prod_{h=t}^{T-1}r(h)\bigg)^{-1},
$$
which deduces that
$$
\pi^*(s,X^{\pi^*}(s),Y^{\pi^*}(s))=\frac{1}{2\mu}\gamma(s)\big[\sigma(s)\sigma(s)^{\top}\big]^{-1}
\bigg(\prod_{h=s}^{T-1}r(h)\bigg)^{-1},t\leq s<T.
$$
Thus, $\pi^*(s,X^{\pi^*}(s),Y^{\pi^*}(s))$ is independent from the states $(X^{\pi^*}(s),Y^{\pi^*}(s))$. The Bellman type dynamic optimal strategy is given as follows:
$$
\pi^*(s)=\frac{1}{2\mu}\gamma(s)\big[\sigma(s)\sigma(s)^{\top}\big]^{-1}
\bigg(\prod_{h=s}^{T-1}r(h)\bigg)^{-1},\ t\leq s<T.
$$
Furthermore, we denote the length of single period as $\displaystyle \frac{T-t}{N}$, and let $N\to\infty$. It follows that
$$
\pi^*(s)=\frac{1}{2\mu}\gamma(s)\big[\sigma(s)\sigma(s)^{\top}\big]^{-1}
e^{\int_s^T[1-r(h)]\mathrm{d}h},
$$
which is consistent with Theorem 3.2 in \cite{Y20}.
\end{remark}
Notice that, $V^{\mu}(t,x,y)$ and $\pi^*(t,x,y)$ are continuous functions of $(x,y)$. Letting $y\to x$, we denote
$$
V^{\mu}(t,x,x)=\lim_{y\to x}V^{\mu}(t,x,y)=-x\prod_{s=t}^{T-1}r(s)-\frac{1}{4\mu}\sum_{s=t}^{T-1}\beta(s),
$$
and related optimal strategy
$$
\pi^*(s)=\frac{1}{2\mu}\gamma(s)\big[\sigma(s)\sigma(s)^{\top}\big]^{-1}
\bigg(\prod_{h=s}^{T-1}r(h)\bigg)^{-1},\ t\leq s<T.
$$
In the following, $\pi^*(\cdot)$ is called the Bellman type dynamic optimal strategy of mean variance model (\ref{cost-1}) under constraints (\ref{mean-1}).

\subsection{Dynamic efficient frontier}
In this section, we derive the dynamic efficient frontier for $\mathbb{E}[X^{{\pi}^{*}}(s)]$ and $\mathrm{Var}[X^{{\pi}^{*}}(s)],\ t\leq s< T$. Plugging the optimal strategy
$$
\pi^*(s)=\frac{1}{2\mu}\gamma(s)\big[\sigma(s)\sigma(s)^{\top}\big]^{-1}
\bigg(\prod_{h=s}^{T-1}r(h)\bigg)^{-1}
, \ t\leq s< T,
$$
into equation (\ref{asset-1}), one can obtain that
$\mathbb{E}[{X}^{{\pi}^{*}}(\cdot)]$ and $\mathbb{E}[\big({X}^{{\pi}^{*}}(\cdot)\big)^2]$ satisfy the following linear difference equations.
\begin{equation}\label{step-asset-1}
\left\{\begin{array}{rl}
\!\!\!\mathbb{E}[X^{{\pi}^{*}}(s)]  & \!\!\!=\displaystyle r(s-1)\mathbb{E}[X^{{\pi}^{*}}(s-1)]
+ \frac{\beta(s-1)}{2\mu}\bigg(\prod_{h=s-1}^{T-1}r(h)\bigg)^{-1},  \\
 \!\!\!\mathbb{E}[X^{{\pi}^{*}}(t)] & \!\!\!=x,\ t< s\leq T,
\end{array}\right.
\end{equation}
and
\begin{equation}\label{step-asset-2}
\left\{\begin{array}{rl}
\!\!\!\mathbb{E}[\big(X^{{\pi}^{*}}(s)\big)^2]  & \!\!\!=\mathbb{E}\bigg[\bigg(\displaystyle r(s-1)X^{{\pi}^{*}}(s-1)
+ \frac{\beta(s-1)}{2\mu}\bigg(\prod_{h=s-1}^{T-1}r(h)\bigg)^{-1}\bigg)^2\bigg]\\
  &+ \displaystyle \frac{\beta(s-1)}{4\mu^2}\bigg(\prod_{h=s-1}^{T-1}r(h)\bigg)^{-2},  \\
 \!\!\!\mathbb{E}[\big(X^{{\pi}^{*}}(t)\big)^2] & \!\!\!=x^2, \ t< s\leq T.
\end{array}\right.
\end{equation}
By equation (\ref{step-asset-1}), we have
\begin{equation}\label{step-asset-1-1}
\left\{\begin{array}{rl}
\!\!\!\big(\mathbb{E}[X^{{\pi}^{*}}(s)]\big)^2  & \!\!\!=\bigg[\displaystyle r(s-1)\mathbb{E}[X^{{\pi}^{*}}(s-1)]
+ \frac{\beta(s-1)}{2\mu}\bigg(\prod_{h=s-1}^{T-1}r(h)\bigg)^{-1}\bigg]^2,  \\
 \!\!\!\mathbb{E}[X^{{\pi}^{*}}(t)]^2 & \!\!\!=x^2,\ t< s\leq T.
\end{array}\right.
\end{equation}
Note that, $\mathrm{Var}[X^{{\pi}^{*}}(s)]=\mathbb{E}[\big(X^{{\pi}^{*}}(s)\big)^2] -\big(\mathbb{E}[X^{{\pi}^{*}}(s)]\big)^2,\ t\leq s\leq T$, combining equations (\ref{step-asset-2}) and (\ref{step-asset-1-1}), it follows that,
\begin{equation}\label{step-asset-3}
\left\{\begin{array}{rl}
\!\!\!\mathrm{Var}[X^{{\pi}^{*}}(s)]  & \!\!\!=\displaystyle
(r(s-1))^2\mathrm{Var}[X^{{\pi}^{*}}(s-1)] +
\frac{\beta(s-1)}{4\mu^2}\bigg(\prod_{h=s-1}^{T-1}r(h)\bigg)^{-2},  \\
 \!\!\!\mathrm{Var}[X^{{\pi}^{*}}(t)] & \!\!\!=0, \ t< s\leq T.
\end{array}\right.
\end{equation}
From equations (\ref{step-asset-1}) and (\ref{step-asset-3}), for $t\leq s\leq T$, we can obtain $\mathbb{E}[X^{{\pi}^{*}}(s)]$ and $\mathrm{Var}[X^{{\pi}^{*}}(s)]$ as follows:
\begin{equation}\label{step-asset-4}
\left\{\begin{array}{rl}
\!\!\!\mathbb{E}[X^{{\pi}^{*}}(s)]&=\displaystyle  x\prod_{h=t}^{s-1}r(h)+\bigg(\prod_{h=s}^{T-1}r(h)\bigg)^{-1}\sum_{h=t}^{s-1}\frac{\beta(h)}{2\mu}                      ,\\
\!\!\!\mathrm{Var}[X^{{\pi}^{*}}(s)]&=\displaystyle  \bigg(\prod_{h=s}^{T-1}r(h)\bigg)^{-2}\sum_{h=t}^{s-1}\frac{\beta(h)}{4\mu^2} .
\end{array}\right.
\end{equation}

\begin{remark}\label{re-2-2}
Notice that, we introduce the risk aversion coefficient $\mu$ in cost functional (\ref{cost-2}). By equation (\ref{step-asset-4}), we can obtain $\mu$ by constrained condition (\ref{mean-1}) as follows:
$$
\displaystyle \mu=\frac{\sum_{h=t}^{T-1}\beta(h)}{2\big(L- x\prod_{h=t}^{T-1}r(h)\big)}.
$$
\end{remark}

From equation (\ref{step-asset-4}), for $t< s\leq T$, the relationship between $\mathbb{E}[X^{{\pi}^{*}}(s)]$ and $\mathrm{Var}[X^{{\pi}^{*}}(s)]$ is given as follows:
\begin{theorem}\label{the-3}
Let Assumptions {$\textbf{H}_1$} and {$\textbf{H}_2$} hold. We have
\begin{equation}\label{eff-1}
\displaystyle \mathrm{Var}[X^{{\pi}^{*}}(s)]=
\frac{\bigg(
\mathbb{E}[X^{{\pi}^{*}}(s)]-x\prod_{h=t}^{s-1}r(h)\bigg)^2}{\sum_{h=t}^{s-1}\beta(h)},\quad
t< s\leq  T,
\end{equation}
where $\beta(h)=\gamma(h)[\sigma(h)\sigma(h)^{\top}]^{-1}\gamma(h)^{\top},\ t\leq h<T$.
\end{theorem}

\subsection{Comparison with pre-committed strategy}

In this part of this paper, we compare our Bellman type dynamic optimal strategy and dynamic efficient frontier with those in \cite{LN00}. Using the same setting and notation of this study, we review the main results of \cite{LN00}, also see \cite{ZL00} for the continuous time case. For the given initial time $t$ and state $x$, the optimal pre-committed strategy is given as follows:
\begin{equation}\label{pre-str-1}
{\pi}^{*}_0(s)=\gamma(s)[\sigma(s)\sigma(s)^{\top}]^{-1}[\lambda \bigg(\prod_{h=s}^{T-1}r(h)\bigg)^{-1} -X^{{\pi}_0^{*}}(s)],\quad  t\leq s< T,
\end{equation}
where $\displaystyle \lambda=\frac{\prod_{h=t}^{T-1}[\beta(h)+1]}{2\mu}+x\prod_{h=t}^{T-1}r(h)$. The related efficient frontier is given as follows:
\begin{equation}\label{pre-eff}
\displaystyle  \mathrm{Var}[X^{{\pi}_0^{*}}(T)]=
\frac{\bigg{(}\mathbb{E}[X^{{\pi}_0^{*}}(T)]
-x\prod_{h=t}^{T-1}r(h)\bigg{)}^2}{\prod_{h=t}^{T-1}[\beta(h)+1]-1},
\end{equation}
where
$$
\displaystyle \mathbb{E}[{X}^{{\pi}^{*}_0}(s)]=x\prod_{h=t}^{s-1}\frac{r(h)}{\beta(h)+1}
+\lambda \bigg(\prod_{h=s}^{T-1}r(h)\bigg)^{-1}[1-\bigg(\prod_{h=t}^{s-1}[\beta(h)+1]\bigg)^{-1}],\ t\leq s\leq T,
$$
and
$$
\displaystyle \mathbb{E}[{X}^{{\pi}^{*}_0}(T)]=
x \prod_{h=t}^{T-1}r(h)+\frac{1}{2\mu}\bigg(\prod_{h=t}^{T-1}[\beta(h)+1]-1\bigg).
$$

Based on our model, by equality (\ref{step-asset-4}), we have
$$
\mathbb{E}[X^{{\pi}^{*}}(s)]=\displaystyle  x\prod_{h=t}^{s-1}r(h)+\bigg(\prod_{h=s}^{T-1}r(h)\bigg)^{-1}\sum_{h=t}^{s-1}\frac{\beta(h)}{2\mu},
$$
with the dynamic optimal strategy
$$
\pi^*(s)=\frac{1}{2\mu}\gamma(s)\big[\sigma(s)\sigma(s)^{\top}\big]^{-1}
\bigg(\prod_{h=s}^{T-1}r(h)\bigg)^{-1}
, \ t\leq s< T,
$$

By formula (\ref{pre-str-1}), the optimal pre-committed strategy $\pi^*_0(\cdot)$ at initial time $t$ is given as follows:
$$
\pi^*_0(t)=\frac{1}{2\mu}\gamma(t)[\sigma(t)\sigma(t)^{\top}]^{-1} \bigg(\prod_{h=t}^{T-1}\frac{r(h)}{\beta(h)+1}\bigg)^{-1}.
$$

 Note that $\beta(\cdot)>0$, we have that $\pi^*(t)<\pi^*_0(t)$, where $\pi^*(t)<\pi^*_0(t)$ means that the absolute value of each element of $\pi^*(t)$ is smaller than that of $\pi^*_0(t)$. This is because the optimal pre-committed strategy cares about the mean and variance of the wealth at investment period $T$, but not the entire investment period $\{t+1,t+2,\cdots,T\}$. Therefore, the optimal pre-committed strategy changes along with the initial time $t$. In contrast to this, our dynamic optimal strategy $\pi^*(\cdot)$ is derived based on minimizing the cost functional along the lines of investment periods $\{t+1,t+2,\cdots,T\}$. Thus, when we provide the dynamic optimal strategy $\pi^*(\cdot)$ at initial time $t$, it will not change in the following periods $s\in\{t+1,t+2,\cdots,T\}$. In the following, we show the properties of mean and variance under the pre-committed strategy $\pi^*_0(\cdot)$ and the dynamic strategy $\pi^*(\cdot)$. The proof of Proposition \ref{pro-1} is given in Appendix \ref{sec:proof}.
\begin{proposition}\label{pro-1}
For a given mean level $L>x\prod_{h=t}^{T-1}r(h)$ at the initial time $t$ under the constrained condition (\ref{mean-1}), we have
\begin{equation}\label{pro-1-var}
\mathrm{Var}[X^{{\pi}^{*}}(T)]>\mathrm{Var}[X^{{\pi}_0^{*}}(T)].
\end{equation}
For a given risk aversion parameter $\mu>0$, we have
\begin{equation}\label{pro-1-var-1}
\mathrm{Var}[X^{{\pi}^{*}}(T)]<\mathrm{Var}[X^{{\pi}_0^{*}}(T)],\quad \mathbb{E}[{X}^{{\pi}^{*}}(T)]<\mathbb{E}[{X}^{{\pi}^{*}_0}(T)].
\end{equation}
\end{proposition}
\begin{remark}
 Note that $L>x\prod_{h=t}^{T-1}r(h)$ at initial time $t$, $\mathbb{E}[{X}^{{\pi}^{*}}(T)]=\mathbb{E}[{X}^{{\pi}^{*}_0}(T)]=L$, based on the dynamic optimal strategy $\pi^*(\cdot)$ and the optimal pre-committed strategy $\pi_0^*(\cdot)$, the variance of the wealth ${X}^{{\pi}^{*}}(T)$ is larger than that of the wealth ${X}^{{\pi}_0^{*}}(T)$. In contrast to this, for a given risk aversion parameter $\mu>0$, the investor can obtain a smaller mean and variance of the wealth ${X}^{{\pi}^{*}}(T)$ at investment period $T$ within the strategy $\pi^*(\cdot)$, relative to the mean and variance of the wealth ${X}^{{\pi}_0^{*}}(T)$ with the strategy $\pi^*_0(\cdot)$. Furthermore, for the given investment period $T$, based on the formulations of $\mathbb{E}[{X}^{{\pi}^{*}}(T)]$ and $\mathbb{E}[{X}^{{\pi}^{*}_0}(T)]$, we can see that the larger risk aversion parameter $\mu$ along with a larger mean level $L$ in constrained condition (\ref{mean-1}).
\end{remark}

\section{Varying investment period mean-variance model}\label{vmvm-1}
Note here, we consider the Bellman type dynamic optimal strategy for the discrete mean-variance model with a given investment period $T$ in Section \ref{mvm-1}. The question is how to determine the investment period $T$. To answer this, we introduce a varying investment period discrete mean-variance model in this section. In the following, we use the notation which is given in Section \ref{mvm-1}.

\subsection{Formulate the model}
In this section, we set the initial time $t=0$ and initial wealth $x>0$. For a given deterministic time $\tau\in \mathbb{N}^+$, we first introduce a varying investment period for the classical discrete mean-variance model:
\begin{equation}
\tau^{\pi}=\inf\{s :\ \mathbb{E}[X^{\pi}(s)]\geq g(\tau),\ 0<s\leq \tau \}\bigwedge \tau,
\end{equation}
where $a_1\bigwedge a_2,\ a_1,a_2\in \mathbb{R}$ means $a_1\bigwedge a_2=\min(a_1,a_2)$, and $g(\cdot)$ is the varying expected return on investment $X^{\pi}(\cdot)$.
\begin{remark}\label{re-3-1}
Note that, $g(\tau)$ is the expected return on investment $X^{\pi}(\cdot)$ before time $\tau$. In this study, we set
\begin{equation}\label{exp-ret}
g(\tau)=x\prod_{h=0}^{\tau-1}r(h)+\alpha x\prod_{h=0}^{\tau-1}\theta(h),\ \tau>0,
\end{equation}
where $x\prod_{h=0}^{\tau-1}r(h)$ is the return when investing all the money into risk-free asset $P_0(\cdot)$, $\alpha x\prod_{h=0}^{\tau-1}\theta(h)$ is the excess return, and $\alpha>0,\ \theta(\cdot)>1$.
\end{remark}
The objective is to minimize the variance at time $\tau^{\pi}$,
\begin{equation}\label{vcost-1}
J(\tau^{\pi},\pi(\cdot))=\mathbb{E}[(X^{\pi}(\tau^{\pi})-\mathbb{E}[X^{\pi}(\tau^{\pi})])^2].
\end{equation}
If there exists $(\bar{\pi}^*(\cdot),\tau^*)$ minimizing the cost functional (\ref{vcost-1}) in the sense of Bellman type time-consistent, we call $\bar{\pi}^*(\cdot)$ the Bellman type dynamic optimal strategy, $\tau^*$ the optimal investment period, and $(\bar{\pi}^*(\cdot),\tau^*)$ the optimal pair.

Note that,
\begin{equation}\label{iter-1}
\inf_{\tau\in \mathbb{N}^+,\pi(\cdot)\in \mathcal{A}_0^{\tau-1}}J(\tau^{\pi},\pi(\cdot))
=\inf_{\tau\in \mathbb{N}^+}
\inf_{\pi(\cdot)\in \mathcal{A}_0^{\tau-1}}J(\tau^{\pi},\pi(\cdot)),
\end{equation}
to obtain the Bellman type dynamic optimal strategy and investment period for the cost functional (\ref{vcost-1}), we give the following steps:

\textbf{Step 1:} For the given $\tau\in \mathbb{N}^+$, we solve the first part $J(\tau^{\bar{\pi}},\bar{\pi}^{\tau} (\cdot))=\inf_{\pi(\cdot)\in \mathcal{A}_0^{\tau-1}}J(\tau^{\pi},\pi(\cdot))$, and obtain the Bellman type optimal strategy $\bar{\pi}^{\tau}(\cdot)$ and $\tau^{\bar{\pi}}=\tau$.

\textbf{Step 2:} We then solve the second part as $J(\tau^*)=\inf_{\tau\in \mathbb{N}^+}J(\tau,\bar{\pi}^{\tau} (\cdot))$, and find the optimal investment period $\tau^*$ and related optimal strategy $\bar{\pi}^{*}(\cdot)$.

\subsection{Solving the mean-variance model}
We first consider the \textbf{Step 1}.  For a given $\tau\in \mathbb{N}^+$, we suppose that there exists an optimal Bellman type strategy $\bar{\pi}^{\tau}(\cdot)$ and investment period $\tau^{\bar{\pi}}\leq \tau$ such that
$$
J(\tau^{\bar{\pi}},\bar{\pi}^{\tau}(\cdot))=\inf_{\pi(\cdot)\in \mathcal{A}_0^{\tau-1}}J(\tau,\pi(\cdot)),
$$
and
$$
\mathbb{E}[X^{\bar{\pi}^{\tau}}(\tau^{\bar{\pi}})]=g(\tau), \ \mathbb{E}[X^{\bar{\pi}^{\tau}}(s)]<g(\tau),\ s<\tau^{\bar{\pi}}.
$$

In the following, we want to show that the optimal strategy which is given in Theorem \ref{the-2} is the strategy $\bar{\pi}^{\tau}(\cdot)$.
\begin{lemma}\label{le-3-1}
Let Assumptions {$\textbf{H}_1$} and {$\textbf{H}_2$} hold. For a given $\tau\in \mathbb{N}^+$, the optimal investment period $\tau^{\bar{\pi}}=\tau$, and the  Bellman type dynamic optimal strategy is
$$
\bar{\pi}^{\tau}(s)=\frac{1}{2\mu}\gamma(s)\big[\sigma(s)\sigma(s)^{\top}\big]^{-1}
\bigg(\prod_{h=s}^{{\tau}-1}r(h)\bigg)^{-1},\ 0\leq s<\tau.
$$
\end{lemma}
\noindent \textbf{Proof:}  Based on the results of Theorem \ref{the-2} and Remark \ref{re-2-1}, for the given investment period $\tau^{\bar{\pi}}$, we can obtain a Bellman type dynamic strategy for the cost functional
$$
J(\tau^{\bar{\pi}},\pi(\cdot))=\mathbb{E}[\big(X^{\pi}(\tau^{\bar{\pi}})
-\mathbb{E}[X^{\pi}(\tau^{\bar{\pi}})]\big)^2]
$$
under the mean constrained
$$
\mathbb{E}[X^{\pi}(\tau^{\bar{\pi}})]=g(\tau).
$$
The optimal strategy is given as follows:
$$
\pi^*(s)=\frac{1}{2\mu}\gamma(s)\big[\sigma(s)\sigma(s)^{\top}\big]^{-1}
\bigg(\prod_{h=s}^{\tau^{\bar{\pi}}-1}r(h)\bigg)^{-1},\ 0\leq s<\tau^{\bar{\pi}}.
$$
By Theorem \ref{the-3}, one can obtain
\begin{equation}\label{eff-2}
\displaystyle \mathrm{Var}[X^{{\pi}^{*}}(s)]=
\frac{\bigg(g(\tau)-x\prod_{h=0}^{s-1}r(h)\bigg)^2}{\sum_{h=0}^{s-1}\beta(h)},
\end{equation}
and
$$
\mathbb{E}[X^{{\pi}^{*}}(s)]=\displaystyle  x\prod_{h=0}^{s-1}r(h)+\bigg(\prod_{h=s}^{\tau^{\bar{\pi}}-1}r(h)\bigg)^{-1}
\sum_{h=0}^{s-1}\frac{\beta(h)}{2\mu},\ s\leq   \tau^{\bar{\pi}}.
$$

Notice that, $\mathbb{E}[X^{{\pi}^{*}}(s)]$ is increasing with $s\leq \tau^{\bar{\pi}}$. Letting
$$
\mu=\frac{\sum_{h=0}^{\tau^{\bar{\pi}}-1}\beta(h)}
{2\big(g(\tau)-x\prod_{h=0}^{\tau^{\bar{\pi}}-1}r(h)\big)},
$$
it follows that
$$
\mathbb{E}[X^{{\pi}^{*}}(\tau^{\bar{\pi}})]=g(\tau),\quad  \mathbb{E}[X^{{\pi}^{*}}(s)]<g(\tau),\ s<\tau^{\bar{\pi}},
$$
and
$$
\tau^{\bar{\pi}}=\inf\{s :\ \mathbb{E}[X^{\pi^*}(s)]\geq g(\tau),\ 0<s\leq \tau \}.
$$
Therefore, we have $\bar{\pi}^{\tau}(\cdot)=\pi^*(\cdot)$.

Now, we determine the varying investment period $\tau^{\bar{\pi}}\leq \tau$. Note that
$$
g(\tau)=x\prod_{h=0}^{\tau-1}r(h)+\alpha x\prod_{h=0}^{\tau-1}\theta(h)>x\prod_{h=0}^{\tau-1}r(h).
$$
By equation (\ref{eff-2}), we can see that  $\mathrm{Var}[X^{{\pi}^{*}}(s)]$ is decreasing with
$ s\leq \tau$. Thus, $\mathrm{Var}[X^{{\pi}^{*}}(\cdot)]$ takes the minimize value at time $\tau$, which deduces the varying investment period $\tau^{\bar{\pi}}=\tau$. This completes the proof. $\qquad  \ \ \  \ \ \ \ \ \ \  \Box $

\bigskip

In the following, we consider the \textbf{Step 2} to find the optimal period $\tau^*$. From the \textbf{Step 1}, we can obtain the Bellman type optimal strategy,
$$
\bar{\pi}^{\tau}(s)=\frac{1}{2\mu}\gamma(s)\big[\sigma(s)\sigma(s)^{\top}\big]^{-1}
\bigg(\prod_{h=s}^{{\tau}-1}r(h)\bigg)^{-1},\ 0\leq s<\tau.
$$
and investment period $\tau^{\bar{\pi}}=\tau$ such that
$$
J(\tau,\bar{\pi}^{\tau}(\cdot))=\inf_{\pi(\cdot)\in \mathcal{A}_0^{\tau-1}}J(\tau^{\pi},\pi(\cdot))
=
\frac{\bigg(g(\tau)-x\prod_{h=0}^{\tau-1}r(h)\bigg)^2}{\sum_{h=0}^{\tau-1}\beta(h)}.
$$

Based on equation (\ref{iter-1}), we now solve the part
$$
\inf_{\tau\in \mathbb{N}^+}J(\tau,\bar{\pi}^{\tau}(\cdot))=\inf_{\tau\in \mathbb{N}^+}
\frac{\bigg(g(\tau)-x\prod_{h=0}^{\tau-1}r(h)\bigg)^2}{\sum_{h=0}^{\tau-1}\beta(h)}.
$$
By Remark \ref{re-3-1}, we have
$$
g(\tau)=x\prod_{h=0}^{\tau-1}r(h)+\alpha x\prod_{h=0}^{\tau-1}\theta(h),
$$
thus,
$$
J(\tau,\bar{\pi}^{\tau}(\cdot))=
\frac{\alpha^2x^2 \bigg(\prod_{h=0}^{\tau-1}\theta(h)\bigg)^2}{\sum_{h=0}^{\tau-1}\beta(h)}.
$$

\begin{theorem}\label{the-4}
Let Assumptions {$\textbf{H}_1$} and {$\textbf{H}_2$} hold, and there exists $\hat{\tau}$ such that
for $s\geq \hat{\tau}$,
$$
(\theta(s)^2-1)\sum_{h=0}^{s-1}\beta(h)-\beta(s)\geq 0.
$$
We can find the optimal investment period $0<\tau^*\leq \hat{\tau}$ such that
$$
J(\tau^*)=\inf_{\tau\in \mathbb{N}^+}J(\tau,\bar{\pi}^{\tau}(\cdot)).
$$
The Bellman type dynamic optimal strategy is
$$
\bar{\pi}^{\tau^*}(s)=\frac{1}{2\mu}\gamma(s)\big[\sigma(s)\sigma(s)^{\top}\big]^{-1}
\bigg(\prod_{h=s}^{{\tau^*}-1}r(h)\bigg)^{-1},\ 0\leq s<\tau^*.
$$
\end{theorem}
\noindent \textbf{Proof:} Based on the formula of $J(\tau,\bar{\pi}^{\tau}(\cdot))$, one obtains
 \begin{equation}
\begin{array}{rl}
&J(\tau+1,\bar{\pi}^{\tau+1}(\cdot))-J(\tau,\bar{\pi}^{\tau}(\cdot))\\
=&
\displaystyle \frac{\alpha^2x^2 \bigg(\prod_{h=0}^{\tau}\theta(h)\bigg)^2}{\sum_{h=0}^{\tau}\beta(h)}-
\frac{\alpha^2x^2 \bigg(\prod_{h=0}^{\tau-1}\theta(h)\bigg)^2}{\sum_{h=0}^{\tau-1}\beta(h)} \\
=&\displaystyle \alpha^2x^2 \bigg(\prod_{h=0}^{\tau-1}\theta(h)\bigg)^2
\frac{(\theta(\tau)^2-1)\sum_{h=0}^{\tau-1}\beta(h)-\beta(\tau)}{\sum_{h=0}^{\tau}\beta(h)
\sum_{h=0}^{\tau-1}\beta(h)}.   \\
\end{array}
\end{equation}
Note that, there exists $\hat{\tau}$ such that
for $s\geq \hat{\tau}$,
$$
(\theta(s)^2-1)\sum_{h=0}^{s-1}\beta(h)-\beta(s)\geq 0,
$$
which indicates that
$$
J(s,\bar{\pi}^{s}(\cdot))-J(\hat{\tau},\bar{\pi}^{\hat{\tau}}(\cdot))\geq 0.
$$
Thus, $J(\tau,\bar{\pi}^{\tau}(\cdot))$ takes the minimize value at $\tau^*$ which satisfies  $\tau^*\leq \hat{\tau}$, and the related dynamic optimal strategy is
$$
\bar{\pi}^{\tau^*}(s)=\frac{1}{2\mu}\gamma(s)\big[\sigma(s)\sigma(s)^{\top}\big]^{-1}
\bigg(\prod_{h=s}^{{\tau^*}-1}r(h)\bigg)^{-1},\ 0\leq s<\tau^*.
$$
This completes the proof. $\qquad  \ \ \  \ \ \ \ \ \ \  \Box $

\begin{example}\label{ex-1} In this example, we consider the Black-Sholes setting. Let $r,b,\sigma,\theta$ be independent from time $t\in \mathbb{N}^+$, $\gamma=(b_1-r,\cdots,b_n-r)$, and $\beta=\gamma [\sigma\sigma^{\top}]^{-1}\gamma^{\top}$. For a given $\tau>0$, the expect return of the wealth $X^{\pi}(\cdot)$ is
$$
g(\tau)=xr^{\tau}+\alpha x\theta^{\tau}.
$$
The cost functional in the \textbf{Step 1} is given as follows:
$$
J(\tau,\bar{\pi}^{\tau}(\cdot))=
\frac{\alpha^2x^2 }{\beta}\frac{\theta^{2\tau}}{\tau},
$$
which deduces 
 \begin{equation}
\begin{array}{rl}
&J(\tau+1,\bar{\pi}^{\tau+1}(\cdot))-J(\tau,\bar{\pi}^{\tau}(\cdot))\\
=&\displaystyle
\frac{\alpha^2x^2 }{\beta}\frac{\theta^{2\tau+2}}{\tau+1}-
\frac{\alpha^2x^2 }{\beta}\frac{\theta^{2\tau}}{\tau}\\
=&\displaystyle
\frac{\alpha^2x^2 \theta^{2\tau}}{\beta}\bigg[\frac{\theta^{2}}{\tau+1}
-\frac{1}{\tau}\bigg]\\
=&\displaystyle
\frac{\alpha^2x^2 \theta^{2\tau}}{\beta}\bigg[\frac{(\theta^{2}-1)\tau-1}{(\tau+1)\tau}
\bigg].
\end{array}
\end{equation}
Note that, $\theta>1$. Let $\displaystyle \hat{\tau}=\lceil\frac{1}{\theta^2-1}\rceil$, we have
$$
(\theta^{2}-1)s-1\geq 0,\quad s\geq \hat{\tau},
$$
and
$$
(\theta^{2}-1)s-1<0,\quad s< \hat{\tau}.
$$
Thus, $J(\tau,\bar{\pi}^{\tau}(\cdot))$ takes the minimize value at $\hat{\tau}$, and $\tau^*=\hat{\tau}$. The related dynamic optimal strategy is
$$
\bar{\pi}^{\tau^*}(s)=\frac{1}{2\mu}\gamma\big[\sigma\sigma^{\top}\big]^{-1}
r^{s-\tau^*},\ 0\leq s<\tau^*.
$$
\end{example}

\section{Simulation and Empirical analysis}\label{ea}
In this section, we want to compare three different strategies in Black-Sholes setting. The expect return for period $\tau$ is given as follows:
$$
g(\tau)=xr^{\tau}+\alpha x\theta^{\tau},\quad \tau\in \mathbb{N}^+.
$$

\textbf{STRATEGY I:} The Bellman type dynamic time-consistent optimal strategy is given in Theorem \ref{the-2} as follows:
$$
\pi_1(s)=\frac{1}{2\mu}\gamma\big[\sigma\sigma^{\top}\big]^{-1}r^{s-\tau},\quad 0\leq s<\tau,
$$
where $\tau$ is the given investment period, and
$$
\mu=\frac{\tau\beta}
{2\big(g(\tau)-xr^{\tau}\big)}=\frac{\tau\beta}{2\alpha x\theta^{\tau} }.
$$
Thus,
$$
\pi_1(s)=x\frac{\alpha \theta^{\tau}}{\tau\beta}\gamma\big[\sigma\sigma^{\top}\big]^{-1}r^{s-\tau},\quad 0\leq s<\tau,
$$
which means that $\pi_1(\cdot)$ is the proportional investment of the initial wealth $x$. The efficient frontier is 
\begin{equation}\label{var-1}
\displaystyle \mathrm{Var}[X^{{\pi}_1}(s)]=
\frac{\bigg(
\mathbb{E}[X^{{\pi}_1}(s)]-xr^s\bigg)^2}{\beta s},\quad
0<s\leq \tau.
\end{equation}

\textbf{STRATEGY II:}
The Bellman type dynamic optimal strategy for varying investment period mean-variance model is given in Theorem \ref{the-4} as follows:
$$
\pi_2(s)=\frac{1}{2\mu}\gamma\big[\sigma\sigma^{\top}\big]^{-1}r^{s-\tau^*},\quad 0\leq s<\tau^*,
$$
where $\displaystyle \tau^*=\lceil\frac{1}{\theta^2-1}\rceil$, and
 $$
\mu=\frac{\tau^*\beta}
{2\big(g(\tau^*)-xr^{\tau^*}\big)}.
$$
We can obtain
$$
\pi_2(s)=x\frac{\alpha \theta^{\tau^*}}{\tau^*\beta}\gamma\big[\sigma\sigma^{\top}\big]^{-1}r^{s-\tau^*},\quad 0\leq s<\tau^*.
$$
Thus, $\pi_2(\cdot)$ is also a proportional investment of the initial wealth $x$. The efficient frontier is
\begin{equation}\label{var-2}
\displaystyle \mathrm{Var}[X^{{\pi}_2}(s)]=
\frac{\bigg(
\mathbb{E}[X^{{\pi}_2}(s)]-xr^s\bigg)^2}{\beta s},\quad
0<s\leq \tau^*.
\end{equation}

\textbf{STRATEGY III:} $1/n$ equality strategy. We assume that there are $n$ kinds of risky assets, and invest $1/n$ of the wealth into each risky asset.
$$
\pi_3(s)=(\frac{X^{\pi_3}(s)}{n},\cdots,\frac{X^{\pi_3}(s)}{n}),\quad 0\leq s<\tau.
$$

\subsection{Simulation results}
In this part, we set the following values of parameters:
$$
r=1.0002,\ b=1.005,\ \theta=1.008, \ d=n=10, \ \alpha=0.5, \ x=1,\ p_0=p_i=1,\ 1\leq i\leq n,
$$
and
\begin{equation}
\sigma_{ij}=\left\{\begin{array}{rl}
0,\ 1\leq i\neq   j\leq n,\\
0.01+0.001i,\ 1\leq i=j\leq n.
\end{array}\right.
\end{equation}

Based on the above value of parameters, we simulate the risk-free asset and risky assets, and compare these three strategies  STRATEGY I, II, and III with the return and variance. Let the length of a single period be one day. For the given initial wealth $x=1$, and investment periods $\tau=30,90$ for STRATEGY I and III, we simulate $M$ times the path of $\{X^{m,\pi_1}(\tau)\}_{m=1}^{M}$ and $\{X^{m,\pi_3}(\tau)\}_{m=1}^{M}$, and define
$$
R_1(\tau)=\frac{\sum_{m=1}^{M}X^{m,\pi_1}(\tau)}{M},\quad
V_1(\tau)=\frac{\sum_{m=1}^{M}\big[X^{m,\pi_1}(\tau)-R_1(\tau)\big]^2}{M}
$$
and
$$
R_3(\tau)=\frac{\sum_{m=1}^{M}X^{m,\pi_3}(\tau)}{M},\quad
V_3(\tau)=\frac{\sum_{m=1}^{M}\big[X^{m,\pi_3}(\tau)-R_3(\tau)\big]^2}{M}.
$$
For the given initial wealth $x=1$, and optimal investment period $\displaystyle \tau^*=\lceil\frac{1}{\theta^2-1}\rceil=63$ for STRATEGY II, we simulate $M$ times the path of $\{X^{m,\pi_2}(\tau^*)\}_{m=1}^{M}$, and define
$$
R_2(\tau^*)=\frac{\sum_{m=1}^{M}X^{m,\pi_2}(\tau^*)}{M},\quad
V_2(\tau^*)=\frac{\sum_{m=1}^{M}\big[X^{m,\pi_2}(\tau^*)-R_2(\tau^*)\big]^2}{M}.
$$
The simulation results are concluded in Table \ref{tab-1}.
\begin{table}[H]
  \centering
  \caption{Simulation results of STRATEGY I, II, and III with different investment period and simulation times $M=5000$}
  \label{tab-1}
  \begin{tabular}{lccccc}
    \toprule
     STRATEGY & Initial wealth & Investment period & Return  &  Variance  \\
    \midrule
    STRATEGY I   & $x=1$ & $\tau=30$    &  $R_1(30)=1.6436$ & $V_1(30)=0.0125$    \\
    STRATEGY I   & $x=1$ & $\tau=90$    &  $R_1(90)=2.0513$ & $V_1(90)=0.0107$    \\
    STRATEGY II  & $x=1$ &$\tau^*=63$   &  $R_2(63)=1.8460$ & $V_2(63)=0.0104$     \\
    STRATEGY III & $x=1$ &$\tau=30$     &  $R_3(30)=1.1617$ & $V_3(30)=0.0010$     \\
    STRATEGY III & $x=1$ &$\tau=63$     &  $R_3(60)=1.3716$ & $V_3(63)=0.0030$     \\
    STRATEGY III & $x=1$ &$\tau=90$     &  $R_3(90)=1.5662$ & $V_3(90)=0.0055$    \\
    \bottomrule
    \hline
  \end{tabular}
\end{table}

 Based on equation (\ref{step-asset-4}), the expected return and variance of STRATEGY I and II are as follows:
\begin{table}[H]
  \centering
  \caption{Theory results of STRATEGY I and II}
  \label{tab-2}
  \begin{tabular}{lccccc}
    \toprule
     STRATEGY & Initial wealth & Investment period & Expected return  &  Expected variance  \\
    \midrule
    STRATEGY I   & $x=1$ & $\tau=30$  &  $g(30)=1.6410$ & $\mathrm{Var}[X^{{\pi}_1}(30)]=0.0126$\\
    STRATEGY I   & $x=1$ & $\tau=90$  &  $g(90)=2.0424$ & $\mathrm{Var}[X^{{\pi}_1}(90)]=0.0109$\\
    STRATEGY II  & $x=1$ &$\tau^*=63$ &  $g(63)=1.8387$ & $\mathrm{Var}[X^{{\pi}_2}(63)]=0.0101$\\
    \bottomrule
    \hline
  \end{tabular}
\end{table}
Comparing the results of Table \ref{tab-1} and \ref{tab-2}, we have that
 \begin{equation}
\begin{array}{rl}
&\left|R_1(30)-g(30) \right|<0.003,\ \ \left|V_1(30)-\mathrm{Var}[X^{{\pi}_1}(30)]\right|<0.0001,\\
&\left|R_1(90)-g(90) \right|<0.009,\ \ \left|V_1(90)-\mathrm{Var}[X^{{\pi}_1}(90)]\right|<0.0002,\\
&\left|R_2(63)-g(63) \right|<0.008,\ \ \left|V_2(63)-\mathrm{Var}[X^{{\pi}_2}(63)]\right|<0.0003,\\
\end{array}
\end{equation}
which verifies that the simulation results coincide with the theory results.

\subsection{Experiment results}\label{sub-ex}

We take the daily data of NASDAQ and Dow Jones from Aug. 03, 2009 to Aug. 02, 2019.
\begin{table}[H]
  \centering
  \caption{Daily data of NASDAQ and Dow Jones}
  \label{tab-3}
  \begin{tabular}{lccccc}
    \toprule
     The index & Initial time & The length (days) & Average value & Standard deviation     \\
    \midrule
   NASDAQ     & Aug. 03, 2009 & 2518 & 4488    & 1798  \\
   Dow Jones  & Aug. 03, 2009 & 2518 & 16890   & 4969 \\
    \bottomrule
    \hline
  \end{tabular}
\end{table}

Employing STRATEGY I, II, and III, we consider investing in risky assets in NASDAQ, Dow Jones and risk-free asset $P_0(\cdot)$. We set the values of  the parameters as follows:
the daily return of $P_0(\cdot)$, $r=1.0002$, the initial wealth $x=1$, the daily excess expected return $\theta=1.008$, and $\alpha=0.5$.

To apply the multi-period investment portfolio model for empirical analysis, we first want to determine the length of a single period and denote a single period's length as $L$ days. For each given $L$, we use the following steps to construct the investment portfolios for STRATEGY I, II, and III: we suppose the prices of index NASDAQ, Dow Jones satisfy the following equation:
$$
P_i(s)=P_i(s-1)\big[b_i+\sigma_i \Delta W(s)\big],\quad 1\leq s, \ i=1,2,
$$
and denote the dataset of index NASDAQ, Dow Jones as $\{P^N(s)\}_{s=1}^{2518}$ and $\{P^D(s)\}_{s=1}^{2518}$, respectively.

\textbf{Step 1:} For a single period's length $L$ and initial time $t$, we firstly give the window of history data $w=m_0L+1$. We use the dataset $\{P^N(s)\}_{s=t-w+1}^{t}$ and $\{P^D(s)\}_{s=t-w+1}^{t}$ to estimate the return $b=(b_1,b_2)$ and volatility $\sigma=(\sigma_1,\sigma_2)$ for STRATEGY I and II at time $t \geq w$. Denoting,
$$
R^N(s)=\frac{P^N(s)-P^N(s-1)}{P^N(s-1)},\quad R^D(s)=\frac{P^D(s)-P^D(s-1)}{P^D(s-1)},\quad t-w+2\leq s\leq t,
$$
and
$$
I^N(t,s)=\sum_{i=1}^LR^N((s-1)L+i+t-w+1),\quad I^D(t,s)=\sum_{i=1}^LR^D((s-1)L+i+t-w+1),\quad 1\leq s\leq m_0.
$$

\textbf{Step 2:} The parameters $b=(b_1,b_2)$ and $\sigma=(\sigma_1,\sigma_2)$ are estimated at time $t$ as
$$
\hat{b}(t)=(\hat{b}_1(t),\hat{b}_2(t))=\bigg(1+\frac{\sum_{s=1}^{m_0}I^N(t,s)}{m_0},
1+\frac{\sum_{s=1}^{m_0}I^D(t,s)}{m_0}\bigg),
$$
and
$$
\hat{\sigma}(t)=(\hat{\sigma}_1(t),\hat{\sigma}_2(t))
=\bigg(\frac{\sum_{s=1}^{m_0}(I^N(t,s)-\hat{b}_1(t)+1)^2}{m_0-1},
\frac{\sum_{s=1}^{m_0}(I^D(t,s)-\hat{b}_2(t)+1)^2}{m_0-1}\bigg).
$$

\textbf{Step 3:} For the given investment period $\tau$, STRATEGY I is given as follows:
$$
\hat{\pi}_1(t,s)=\bigg(\frac{r_L^{s-\tau}(\hat{b}_1(t)-r_L)}{2\hat{\mu}_1(t)\hat{\sigma}^2_1(t)},
\frac{r_L^{s-\tau}(\hat{b}_2(t)-r_L)}{2\hat{\mu}_1(t)\hat{\sigma}^2_2(t)}\bigg),\ 0\leq s<\tau,
$$
where $r_L=1+(r-1)L$, and
$$
\hat{\mu}_1(t)=\frac{\tau\hat{\beta}(t)}
{2\big(g(\tau)-xr_L^{\tau}\big)},\quad  \hat{\beta}(t)=\frac{(\hat{b}_1(t)-r_L)^2}{\hat{\sigma}^2_1(t)}
+\frac{(\hat{b}_2(t)-r_L)^2}{\hat{\sigma}^2_2(t)}.
$$
Applying STRATEGY I, the wealth is
 \begin{equation}
\left\{\begin{array}{rl}
\bar{X}^{\hat{\pi}_1}(t,s)=&\displaystyle\frac{r_L^{s-\tau}(\hat{b}_1(t)-r_L)}{2\hat{\mu}_1(t)\hat{\sigma}^2_1(t)}\frac{P^N(t+sL)}{P^N(t+(s-1)L)}
                         +\displaystyle\frac{r_L^{s-\tau}(\hat{b}_2(t)-r_L)}{2\hat{\mu}_1(t)\hat{\sigma}^2_2(t)}\frac{P^D(t+sL)}{P^D(t+(s-1)L)}\\
 &\displaystyle+\bigg(\bar{X}^{\hat{\pi}_1}(t,s-1)-\frac{r_L^{s-\tau}(\hat{b}_1(t)-r_L)}{2\hat{\mu}_1(t)\hat{\sigma}^2_1(t)}
 -\frac{r_L^{s-\tau}(\hat{b}_2(t)-r_L)}{2\hat{\mu}_1(t)\hat{\sigma}^2_2(t)}\bigg)r_L,\\
 \bar{X}^{\hat{\pi}_1}(t,0)=&1,\quad 1\leq s\leq \tau.
 \end{array}\right.
\end{equation}

The optimal investment period of STRATEGY II is $\displaystyle \tau^*=\lceil\frac{1}{\big((\theta-1)L+1\big)^2-1}\rceil$, and STRATEGY II is given as follows:
$$
\hat{\pi}_2(s)=\bigg(\frac{r_L^{s-\tau^*}(\hat{b}_1(t)-r_L)}{2\hat{\mu}_2(t)\hat{\sigma}^2_1(t)},
\frac{r_L^{s-\tau^*}(\hat{b}_2(t)-r_L)}{2\hat{\mu}_2(t)\hat{\sigma}^2_2(t)}\bigg),\ 0\leq s<\tau^*,
$$
and
$$
\hat{\mu}_2(t)=\frac{\tau\hat{\beta}(t)}
{2\big(g(\tau^*)-xr_L^{\tau^*}\big)},\quad \hat{\beta}(t)=\frac{(\hat{b}_1(t)-r_L)^2}{\hat{\sigma}^2_1(t)}
+\frac{(\hat{b}_2(t)-r_L)^2}{\hat{\sigma}^2_2(t)}.
$$
Applying STRATEGY II, the wealth is
 \begin{equation}
\left\{\begin{array}{rl}
\bar{X}^{\hat{\pi}_2}(t,s)=&\displaystyle\frac{r_L^{s-\tau^*}(\hat{b}_1(t)-r_L)}{2\hat{\mu}_2(t)\hat{\sigma}^2_1(t)}\frac{P^N(t+sL)}{P^N(t+(s-1)L)}
                         +\displaystyle\frac{r_L^{s-\tau^*}(\hat{b}_2(t)-r_L)}{2\hat{\mu}_2(t)\hat{\sigma}^2_2(t)}\frac{P^D(t+sL)}{P^D(t+(s-1)L)}\\
 &\displaystyle+\bigg(\bar{X}^{\hat{\pi}_2}(t,s-1)-\frac{r_L^{s-\tau^*}(\hat{b}_1(t)-r_L)}{2\hat{\mu}_2(t)\hat{\sigma}^2_1(t)}
 -\frac{r_L^{s-\tau^*}(\hat{b}_2(t)-r_L)}{2\hat{\mu}_2(t)\hat{\sigma}^2_2(t)}\bigg)
 r_L,\\
 \bar{X}^{\hat{\pi}_2}(t,0)=&1,\quad 1\leq s\leq \tau^*.
 \end{array}\right.
\end{equation}

For the investment period $\tau$, the STRATEGY III is given as follows:
$$
\hat{\pi}_3(s)=\bigg(\displaystyle \frac{\bar{X}^{\hat{\pi}_3}(t,s)}{2},
\frac{\bar{X}^{\hat{\pi}_3}(t,s)}{2}\bigg),\quad 0\leq s<\tau^*,
$$
and the wealth as of STRATEGY III is
 \begin{equation}
\left\{\begin{array}{rl}
\bar{X}^{\hat{\pi}_3}(t,s)=&\displaystyle \frac{\bar{X}^{\hat{\pi}_3}(t,s-1)}{2}\frac{P^N(t+sL)}{P^N(t+(s-1)L)}
+\frac{\bar{X}^{\hat{\pi}_3}(t,s-1)}{2}\frac{P^D(t+sL)}{P^D(t+(s-1)L)},\\
 \bar{X}^{\hat{\pi}_3}(t,0)=&1,\quad 1\leq s\leq \tau.
 \end{array}\right.
\end{equation}

\textbf{Step 4:} For the given initial time $t$, the length of a single period $L$ and the investment period $\tau,\ \tau^*$. We repeat the \textbf{Steps 1, 2, 3} for $w\leq t\leq w+K-1$, and denotes the average yearly return and Sharpe ratio for STRATEGY I, II and III at investment period $\tau,\ \tau^*$ as follows:
\begin{equation*}
\begin{array}{rl}
& \displaystyle \mathrm{Return}_1(L,\tau)=\frac{250}{\tau LK}\sum_{i=1}^{K}(\bar{X}^{\hat{\pi}_1}(w+i-1,\tau)-1), \\
&\mathrm{Sharpe}_1(L,\tau)=\displaystyle
\sqrt{\frac{250}{\tau LK}}\frac{\sum_{i=1}^{K}\bar{X}^{\hat{\pi}_1}(w+i-1,\tau)-K-0.0002\tau L K}
{\sqrt{\sum_{i=1}^{K}\big[\bar{X}^{\hat{\pi}_1}(w+i-1,\tau)-\frac{1}{K}\sum_{i=1}^{K}\bar{X}^{\hat{\pi}_1}(w+i-1,\tau)\big]^2}},                            \\
& \displaystyle \mathrm{Return}_2(L,\tau^*)=\frac{250}{\tau^* LK}\sum_{i=1}^{K}(\bar{X}^{\hat{\pi}_2}(w+i-1,\tau^*)-1), \\
&\mathrm{Sharpe}_2(L,\tau^*)=\displaystyle
\sqrt{\frac{250}{\tau^* LK}}\frac{\sum_{i=1}^{K}\bar{X}^{\hat{\pi}_2}(w+i-1,\tau^*)-K-0.0002\tau^* L K}
{\sqrt{\sum_{i=1}^{K}\big[\bar{X}^{\hat{\pi}_2}(w+i-1,\tau^*)-\frac{1}{K}\sum_{i=1}^{K}\bar{X}^{\hat{\pi}_2}(w+i-1,\tau^*)\big]^2}},                            \\
& \displaystyle \mathrm{Return}_3(L,\tau)=\frac{250}{\tau LK}\sum_{i=1}^{K}(\bar{X}^{\hat{\pi}_3}(w+i-1,\tau)-1), \\
&\mathrm{Sharpe}_3(L,\tau)=\displaystyle
\sqrt{\frac{250}{\tau LK}}\frac{\sum_{i=1}^{K}\bar{X}^{\hat{\pi}_3}(w+i-1,\tau)-K-0.0002\tau L K}
{\sqrt{\sum_{i=1}^{K}\big[\bar{X}^{\hat{\pi}_3}(w+i-1,\tau)-\frac{1}{K}\sum_{i=1}^{K}\bar{X}^{\hat{\pi}_3}(w+i-1,\tau)\big]^2}}.                           \\
\end{array}
\end{equation*}

\bigskip

Based on the above steps, we now show how to choose the length of single period $L$. In the following, we set $m_0=20$, $K=1000$, $\displaystyle \tau=\lceil\frac{250}{L}\rceil$, $\displaystyle \tau^*=\lceil\frac{1}{\big((\theta-1)L+1\big)^2-1}\rceil$ and plot the value of return $\mathrm{Return}_1(L,\tau),\ \mathrm{Return}_2(L,\tau^*),\ \mathrm{Return}_3(L,\tau)$ and
$\mathrm{Sharpe}_1(L,\tau),\ \mathrm{Sharpe}_2(L,\tau^*)$, $\mathrm{Sharpe}_3(L,\tau)$ for $1\leq L \leq 60$:
\begin{figure}[H]
 \caption{Average return and Sharpe ratio of STRATEGY I, II, and III along with the single period $L$}
 \label{fig1}
\begin{center}
\includegraphics[width=3.0 in]{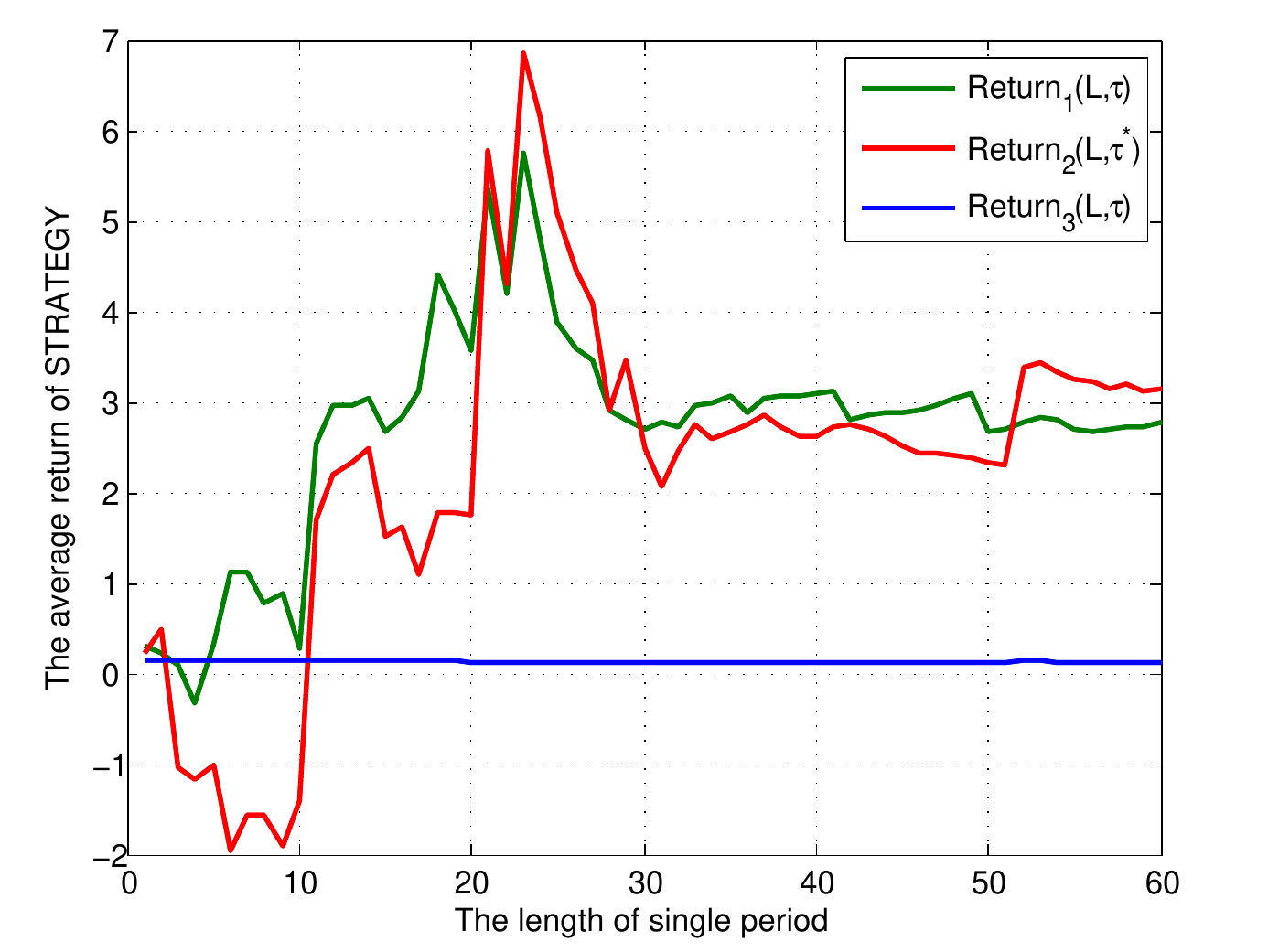}
\includegraphics[width=3.05 in]{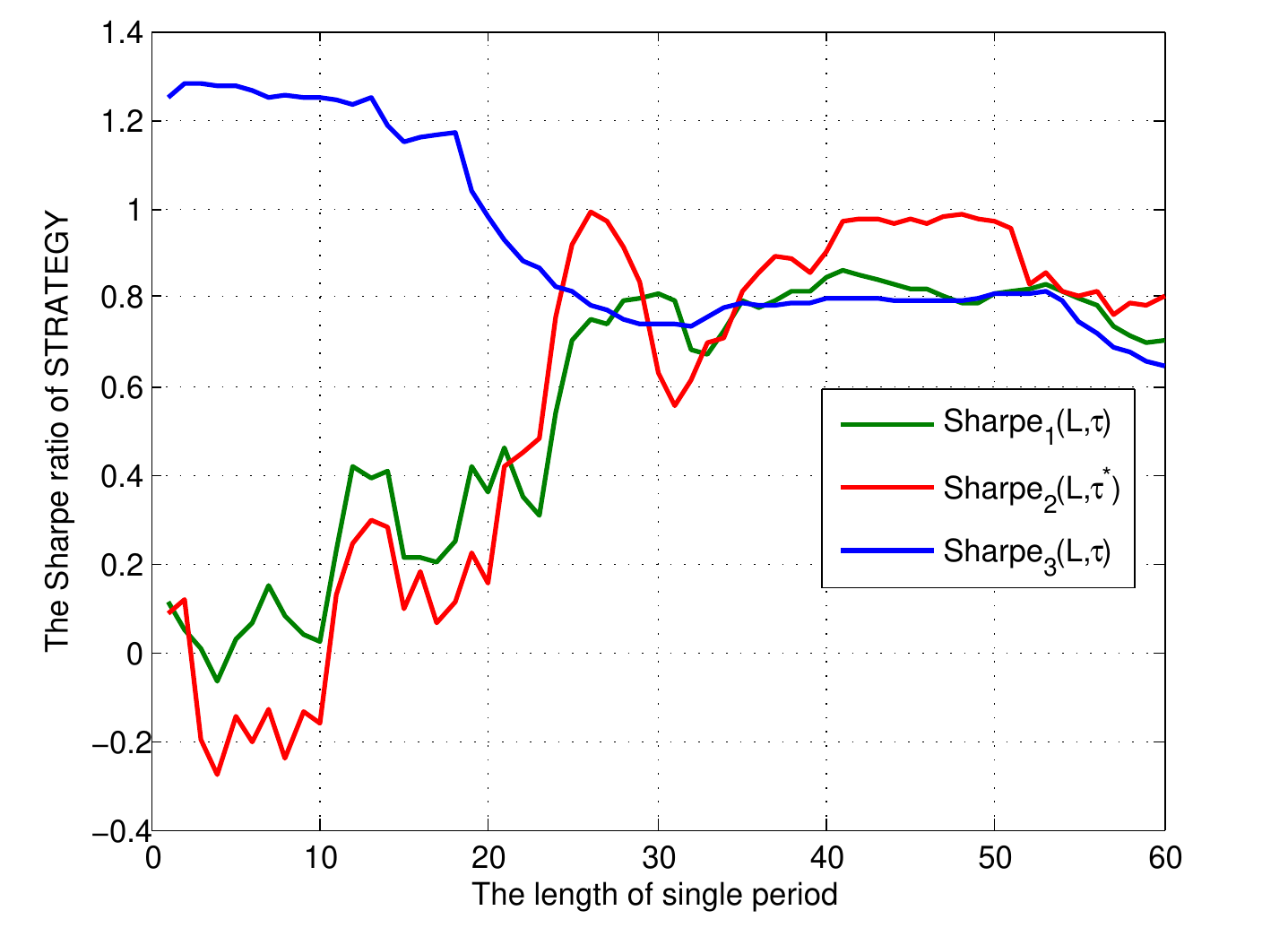}
\end{center}
\end{figure}
From Figure \ref{fig1}, we can see that the return of STRATEGY I is almost equal to that of STRATEGY II when the length of single period $L\geq 10$, and the return of STRATEGY I is almost larger than that of STRATEGY III for $L\geq 1$. Furthermore, when $L\geq 30$, the returns of STRATEGY I and STRATEGY II are stable. However, the return of STRATEGY III barely changes along with the
single period $L\geq 1$. Thus, we plot the Sharpe ratio of STRATEGY I, II and III in Figure \ref{fig1}. We can see that the Sharpe ratio of STRATEGY I is almost same as that of STRATEGY II when the length of single-period $L\geq 10$, and the Sharpe ratio of STRATEGY III is almost equal with that of STRATEGY I and II when the length of single-period $L\geq 30$. Combining the results in Figure \ref{fig1}, we show the results in details for the single-period $L=30$, and investment periods $\tau=9,\ \tau^*=2$.
\begin{figure}[H]
 \caption{Repeat the investment portfolio of STRATEGY I, II, and III 1000 times}
 \label{fig3}
\begin{center}
\includegraphics[width=3.0 in]{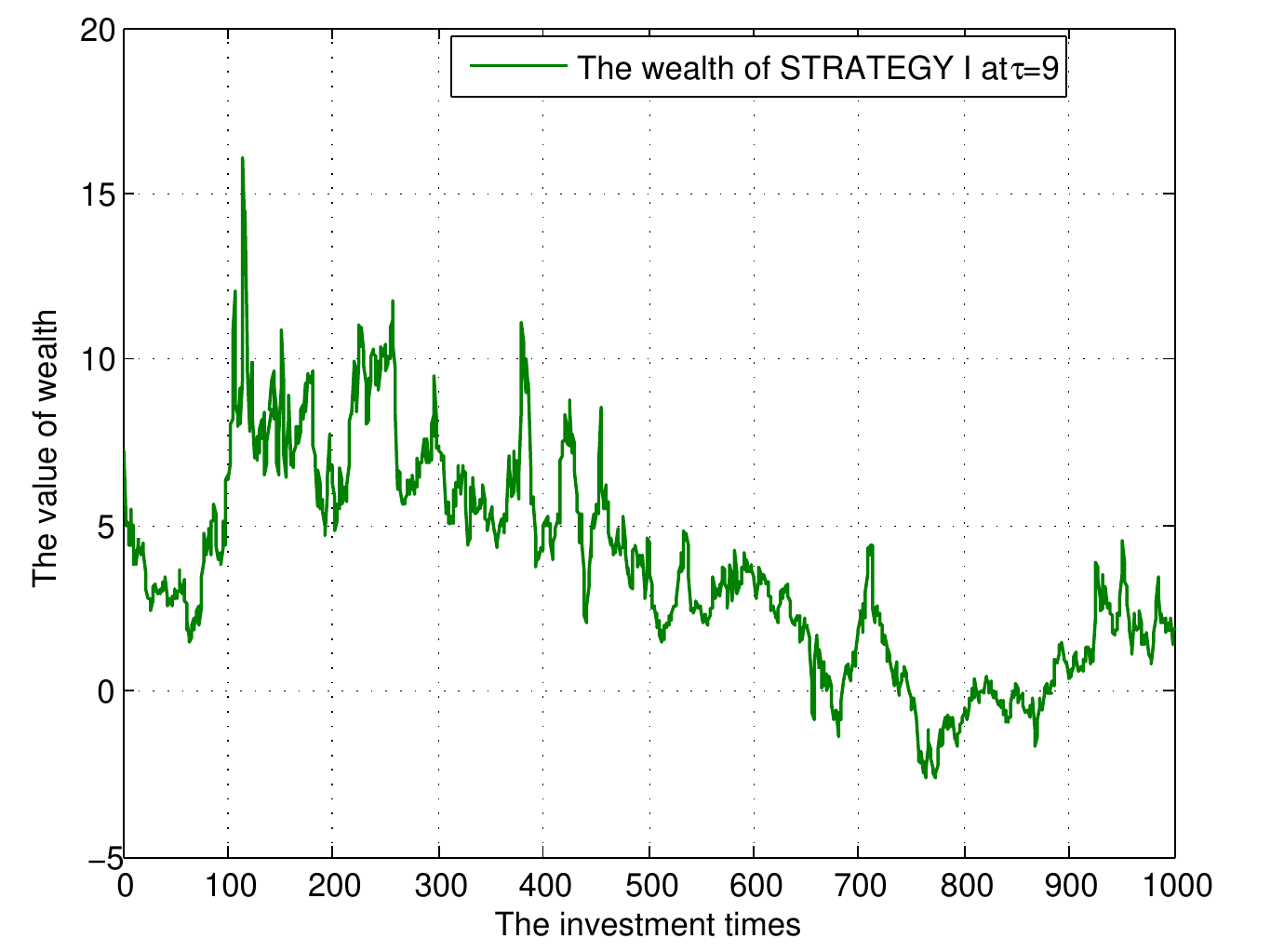}
\includegraphics[width=3.0 in]{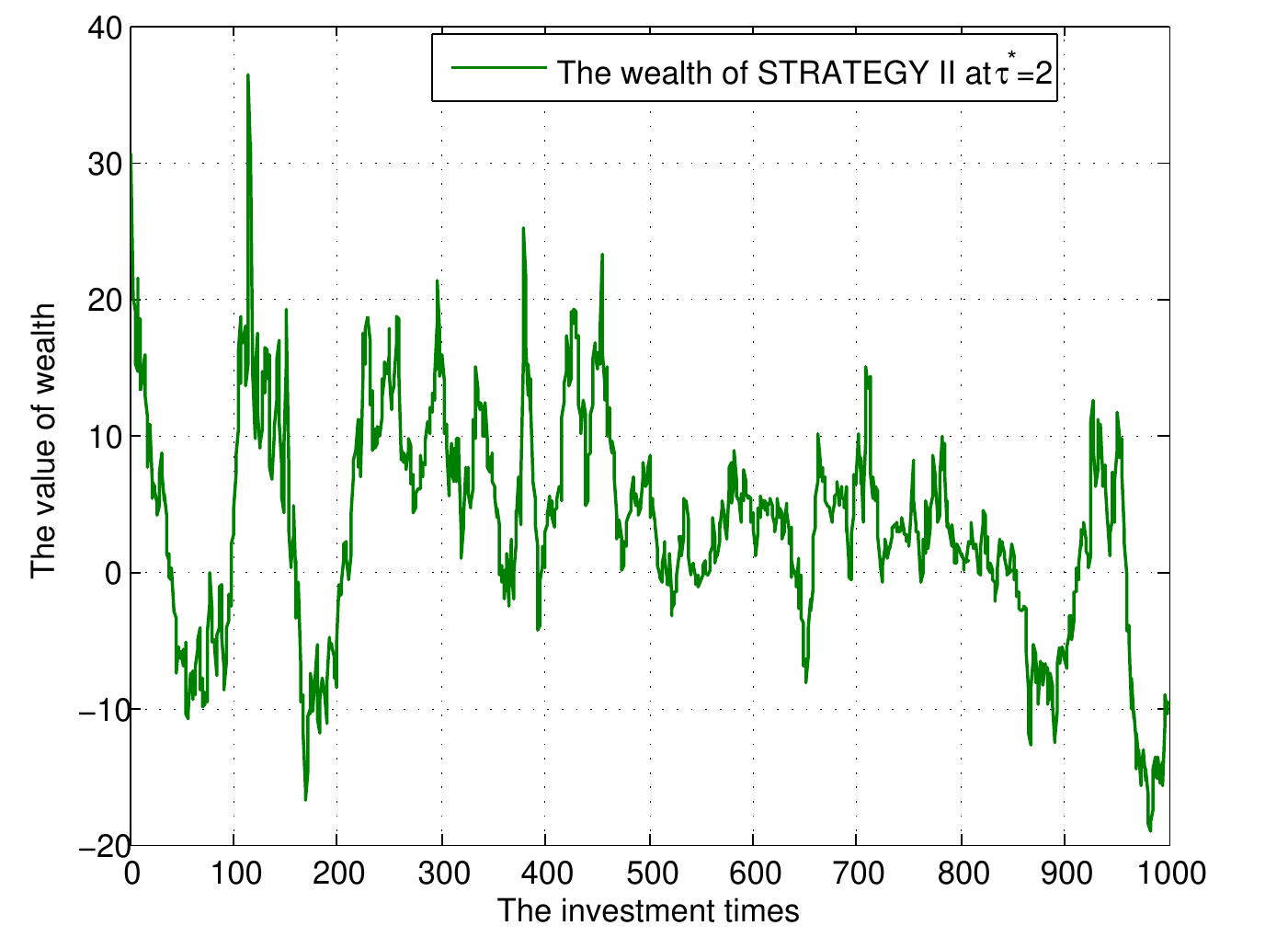}
\includegraphics[width=3.0 in]{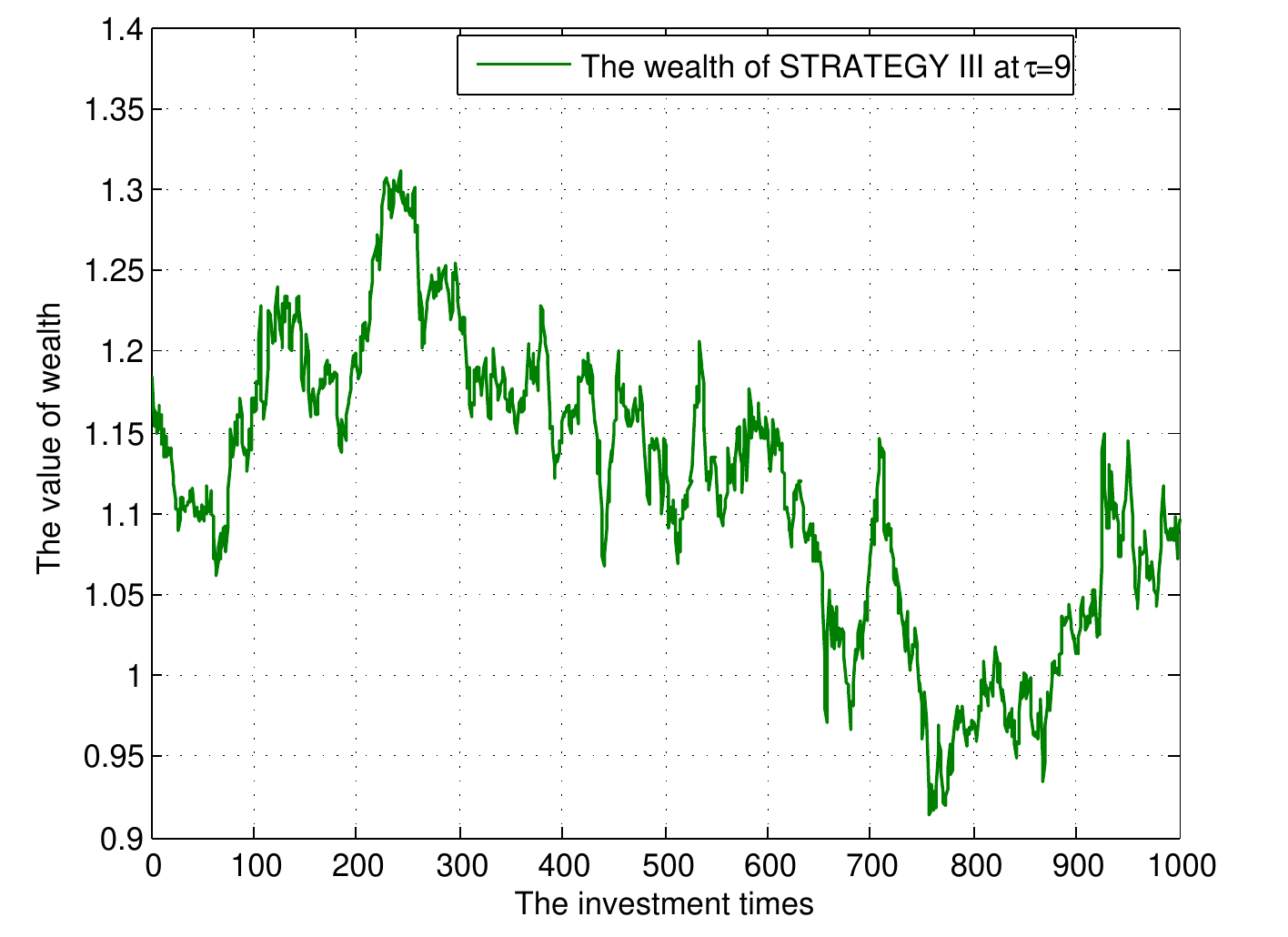}
\includegraphics[width=3.0 in]{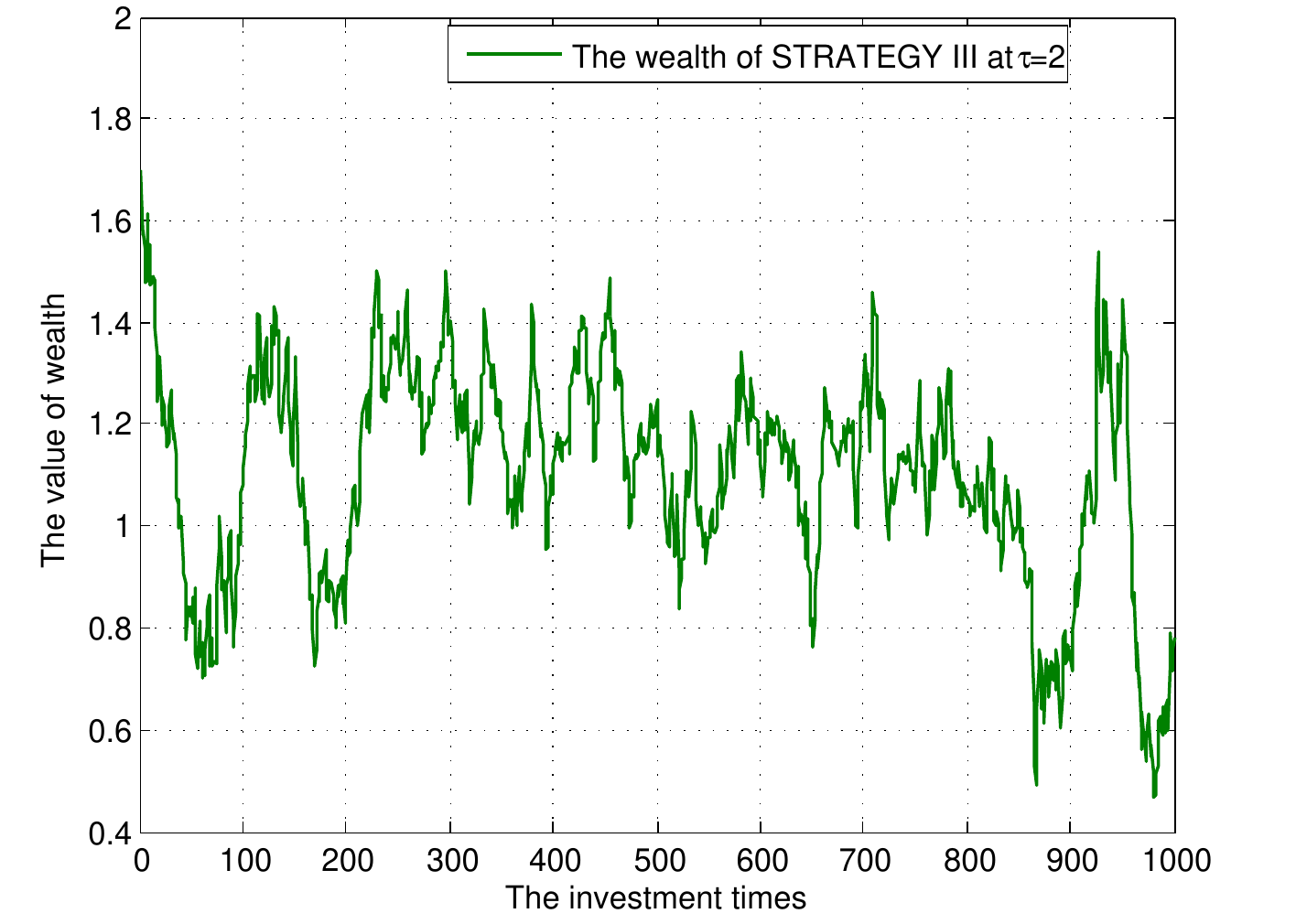}
\end{center}
\end{figure}
In Figure \ref{fig3}, we repeat the multi-period portfolio investment of STRATEGY I, II, and III 1000 times along with $w\leq t\leq w+999$. We conclude that the average yearly return ${\mathrm{Return}}_i(\tau,L),\ i=1,2,3$ and Sharpe ratio ${\mathrm{Sharpe}}_i(\tau,L),\ i=1,2,3$ of STRATEGY I, II, and III are as follows:
\begin{table}[H]
  \centering
  \caption{Empirical analysis of STRATEGY I, II, and III}
  \label{tab-4}
  \begin{tabular}{lccccc}
    \toprule
           & Single period's length  & Investment period   & Yearly return & Sharpe ratio   \\
    \midrule
   STRATEGY I   & L=30  & $\tau=9$  & 2.7078  & 0.8077   \\
   STRATEGY II  & L=30  & $\tau^*=2$ & 2.4912  & 0.6287 \\
   STRATEGY III  & L=30  & $\tau=2$ & 0.1047  & 0.5334 \\
   STRATEGY III & L=30  & $\tau=9$  & 0.1165  & 0.7370  \\
    \bottomrule
    \hline
  \end{tabular}
\end{table}
In Table \ref{tab-4}, we can see that the average yearly return of STRATEGY I is almost equal to that of STRATEGY II and is larger than that of STRATEGY III. The Sharpe ratio of STRATEGY I is almost equal to that of STRATEGY III, and is larger than that of STRATEGY II. Thus, the STRATEGY I is better than the other two strategies based on the length of single period $L=30$. Note that, the STRATEGY I and II have same formula but with different investment periods, $\tau=9>\tau^*=2$. These results indicate that the multi-period investment portfolio strategy maybe better than the single-period investment portfolio strategy.

Based on the results of theory, the expected yearly return $\overline{\mathrm{Return}}_i(\tau,L),\ i=1,2$ and Sharpe ratio $\overline{\mathrm{Sharpe}}_i(\tau,L),\ i=1,2$ for STRATEGY I and II are given as follows:
\begin{equation*}
\begin{array}{rl}
& \displaystyle \overline{\mathrm{Return}}_1(\tau,L)=\frac{250}{L\tau}(g(\tau)-x),\\
& \displaystyle \overline{\mathrm{Sharpe}}_1(\tau,L)=\sqrt{\frac{250}{L\tau}}\frac{K(g(\tau)-x-0.0002\tau L )}{\sum_{t=w}^{w+K-1}\frac{\alpha x (1+(\theta-1)L)^{\tau}}{\sqrt{\tau \hat{\beta}(t)}}},\\
 & \displaystyle \overline{\mathrm{Return}}_2(\tau^*,L)=\frac{250}{L\tau^*}(g(\tau^*)-x),\\
& \displaystyle \overline{\mathrm{Sharpe}}_2(\tau^*,L)=\sqrt{\frac{250}{L\tau^*}}\frac{K(g(\tau)-x-0.0002\tau^* L )}{\sum_{t=w}^{w+K-1}\frac{\alpha x (1+(\theta-1)L)^{\tau^*}}{\sqrt{\tau^* \hat{\beta}(t)}}},\\
\end{array}
\end{equation*}
where $\hat{\beta}(t)$ is calculated at time $t$,
$$
\hat{\beta}(t)=\frac{(\hat{b}_1(t)-r_L)^2}{\hat{\sigma}^2_1(t)}
+\frac{(\hat{b}_2(t)-r_L)^2}{\hat{\sigma}^2_2(t)},\quad w\leq t\leq w+K-1.
$$

\begin{table}[H]
  \centering
  \caption{Theory results of STRATEGY I and II}
  \label{tab-5}
  \begin{tabular}{lccccc}
    \toprule
                 & Investment period & Expected yearly return  &  Sharpe ratio  \\
    \midrule
    STRATEGY I    & $\tau=9$  &  $3.2600 $ & $0.9605$\\
    STRATEGY II   &$\tau^*=2$ &  $3.2535$  & $0.9602$\\
    \bottomrule
    \hline
  \end{tabular}
\end{table}
Comparing Tables \ref{tab-4} and \ref{tab-5}, we can see that,
 \begin{equation}
\begin{array}{rl}
&\left|\mathrm{Return}_1(9,30)-\overline{\mathrm{Return}}_1(9,30) \right|<0.56,\ \ \left|\mathrm{Sharpe}_1(9,30)-\overline{\mathrm{Sharpe}}_1(9,30)\right|<0.16,\\
&\left|\mathrm{Return}_2(2,30)-\overline{\mathrm{Return}}_2(2,30) \right|<0.77,\ \ \left|\mathrm{Sharpe}_2(2,30)-\overline{\mathrm{Sharpe}}_2(2,30)\right|<0.34,\\
\end{array}
\end{equation}
which indicates that the experiment results of STRATEGY I and II are nearly with the theory results.

\subsection{STRATEGY I with transaction fee and loan interest rate}
In this part, we introduce transaction fee rate $r_0$ of risky assets and daily loan interest rate $\overline{r}$ for the discrete time multi-period mean-variance model. We show the details of STRATEGY I with the length of single period $L$ and investment period $\tau$. Using the same setting in Subsection \ref{sub-ex}, we consider to invest into risky assets NASDAQ, Dow Jones and risk-free asset $P_0(\cdot)$. We set the values of  the parameters as follows:
the daily return of risk-free asset, $r=1.0002$, the initial wealth $x=1$, the daily excess expected return $\theta=1.008$, $\alpha=0.5$, the transaction fee rate $r_0=0.001$  and the loan interest rate $\overline{r}=1.0003$.

For the given investment period $\tau$, STRATEGY I is given as follows:
$$
\hat{\pi}_1(t,s)=\bigg(\frac{r_L^{s-\tau}(\hat{b}_1(t)-r_L)}{2\hat{\mu}_1(t)\hat{\sigma}^2_1(t)},
\frac{r_L^{s-\tau}(\hat{b}_2(t)-r_L)}{2\hat{\mu}_1(t)\hat{\sigma}^2_2(t)}\bigg),\ 0\leq s<\tau,
$$
where $r_L=1+(r-1)L$ and $\tau$ is the given investment period, and
$$
\hat{\mu}_1(t)=\frac{\tau\hat{\beta}(t)}
{2\big(g(\tau)-xr_L^{\tau}\big)},\quad \hat{\beta}(t)=\frac{(\hat{b}_1(t)-r_L)^2}{\hat{\sigma}^2_1(t)}
+\frac{(\hat{b}_2(t)-r_L)^2}{\hat{\sigma}^2_2(t)}.
$$

Denoting
\begin{equation}
\left\{\begin{array}{rl}
&\omega_1(t,s)=\displaystyle\frac{r_L^{s-\tau}(\hat{b}_1(t)-r_L)}{2\hat{\mu}_1(t)\hat{\sigma}^2_1(t)},\quad 1\leq s\leq \tau,\\
&\omega_2(t,s)=\displaystyle\frac{r_L^{s-\tau}(\hat{b}_2(t)-r_L)}{2\hat{\mu}_1(t)\hat{\sigma}^2_2(t)}), \quad 1\leq s\leq \tau,\\
&\omega_3(t,s)=\displaystyle\bar{X}^{\hat{\pi}_1}(t,s-1)-\frac{r_L^{s-\tau}(\hat{b}_1(t)-r_L)}{2\hat{\mu}_1(t)\hat{\sigma}^2_1(t)}
 -\frac{r_L^{s-\tau}(\hat{b}_2(t)-r_L)}{2\hat{\mu}_1(t)\hat{\sigma}^2_2(t)}, \quad 1\leq s\leq \tau.
\end{array}\right.
\end{equation}
Based on STRATEGY I, the related wealth is
 \begin{equation}
\left\{\begin{array}{rl}
\bar{X}^{\hat{\pi}_1}(t,s)=&\displaystyle \omega_1(t,s)\frac{P^N(t+sL)}{P^N(t+(s-1)L)}
                         +\omega_2(t,s)\frac{P^D(t+sL)}{P^D(t+(s-1)L)}\\
                         &+\max(\omega_3(t,s),0)r_L
                         +\min(\omega_3(t,s),0)\overline{r}_L-(\left|\omega_1(t,s)\right|
                         +\left|\omega_2(t,s)\right|)r_0\\
 \bar{X}^{\hat{\pi}_1}(t,0)=&1,\quad 1\leq s\leq \tau,
 \end{array}\right.
\end{equation}
where $\overline{r}_L=1+(\overline{r}-1)L$.

Based on the introduction of the transaction fee rate of stocks and loan interest rate, we show the relation between the length of single period $L$ and the return and Sharpe ratio of STRATEGY I. In the following, we set $m_0=20$, $K=1000$, $\displaystyle \tau=\lceil\frac{250}{L}\rceil$,  and plot the value of average return $\mathrm{Return}_1(L,\tau)$ and Sharpe ratio
$\mathrm{Sharpe}_1(L,\tau)$ for $1\leq L \leq 60$:
where \begin{equation*}
\begin{array}{rl}
& \displaystyle \mathrm{Return}_1(L,\tau)=\frac{250}{\tau LK}\sum_{i=1}^{K}(\bar{X}^{\hat{\pi}_1}(w+i-1,\tau)-1), \\
&\mathrm{Sharpe}_1(L,\tau)=\displaystyle
\sqrt{\frac{250}{\tau LK}}\frac{\sum_{i=1}^{K}\bar{X}^{\hat{\pi}_1}(w+i-1,\tau)-K-0.0002\tau L K}
{\sqrt{\sum_{i=1}^{K}\big[\bar{X}^{\hat{\pi}_1}(w+i-1,\tau)-\frac{1}{K}\sum_{i=1}^{K}\bar{X}^{\hat{\pi}_1}(w+i-1,\tau)\big]^2}}.                            \\
\end{array}
\end{equation*}

\begin{figure}[H]
 \caption{Average return and Sharpe ratio of STRATEGY I along with the single-period $L$}
 \label{fig4}
\begin{center}
\includegraphics[width=3.0 in]{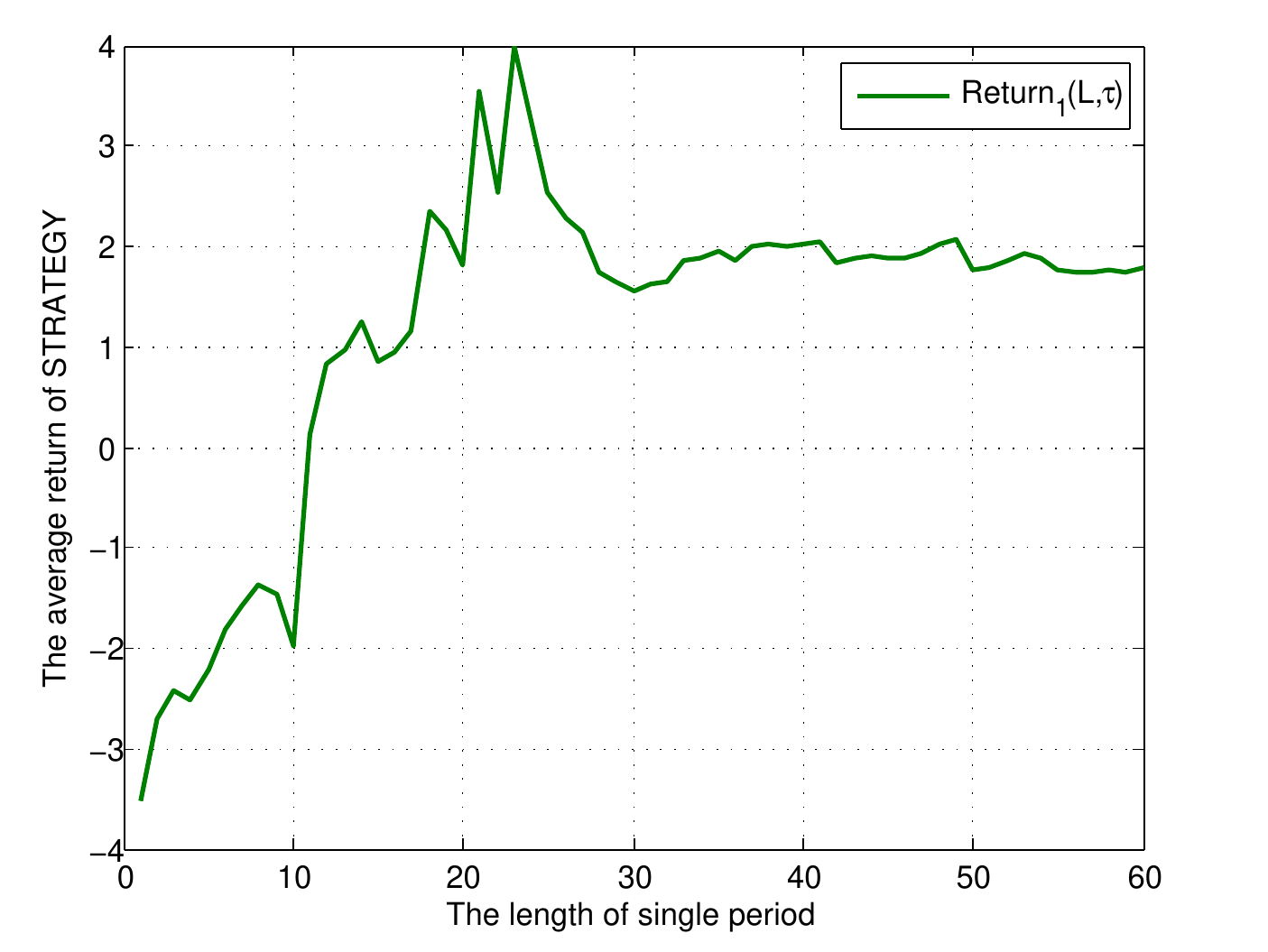}
\includegraphics[width=3.05 in]{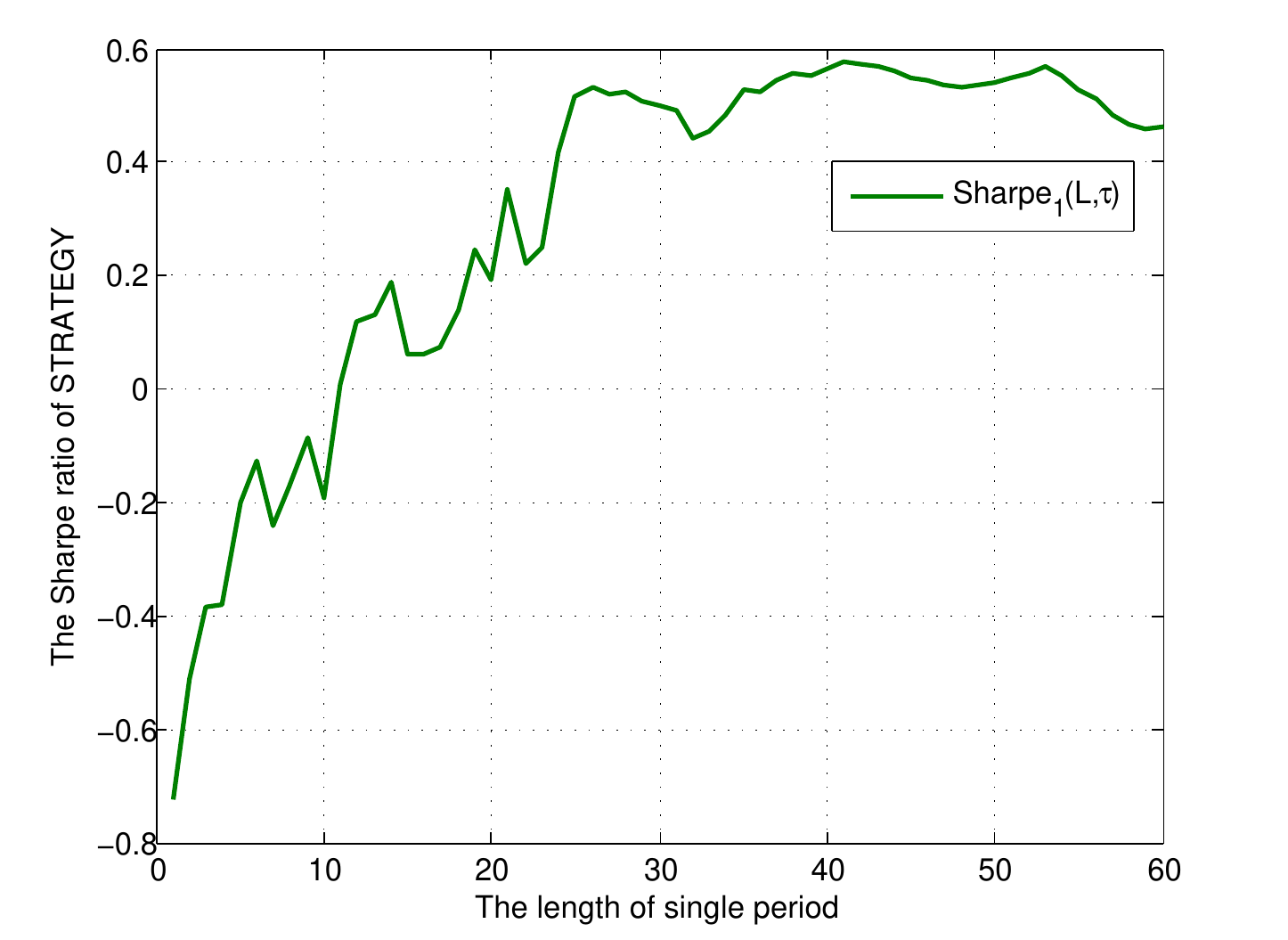}
\end{center}
\end{figure}

In Figure \ref{fig4}, we build the investment portfolio for NASDAQ, Dow Jones, and risk-free asset using STRATEGY I.
After considering the transaction fee rate $r_0$ for risky assets and loan interest rate $\overline{r}_L$, we find that the average return of the portfolio of STRATEGY I is larger than $0$ when the length of single period $L> 10$ and becomes stable when $L\geq 30$. Furthermore, we can see that the Sharpe ratio of STRATEGY I is nearly $0.6$ when $L\geq 30$. Therefore, we consider the details of investment portfolio of STRATEGY I when $L=30$.

\begin{figure}[H]
 \caption{Repeat the investment portfolio of STRATEGY I 1500 times }
 \label{fig5}
\begin{center}
\includegraphics[width=3.8 in]{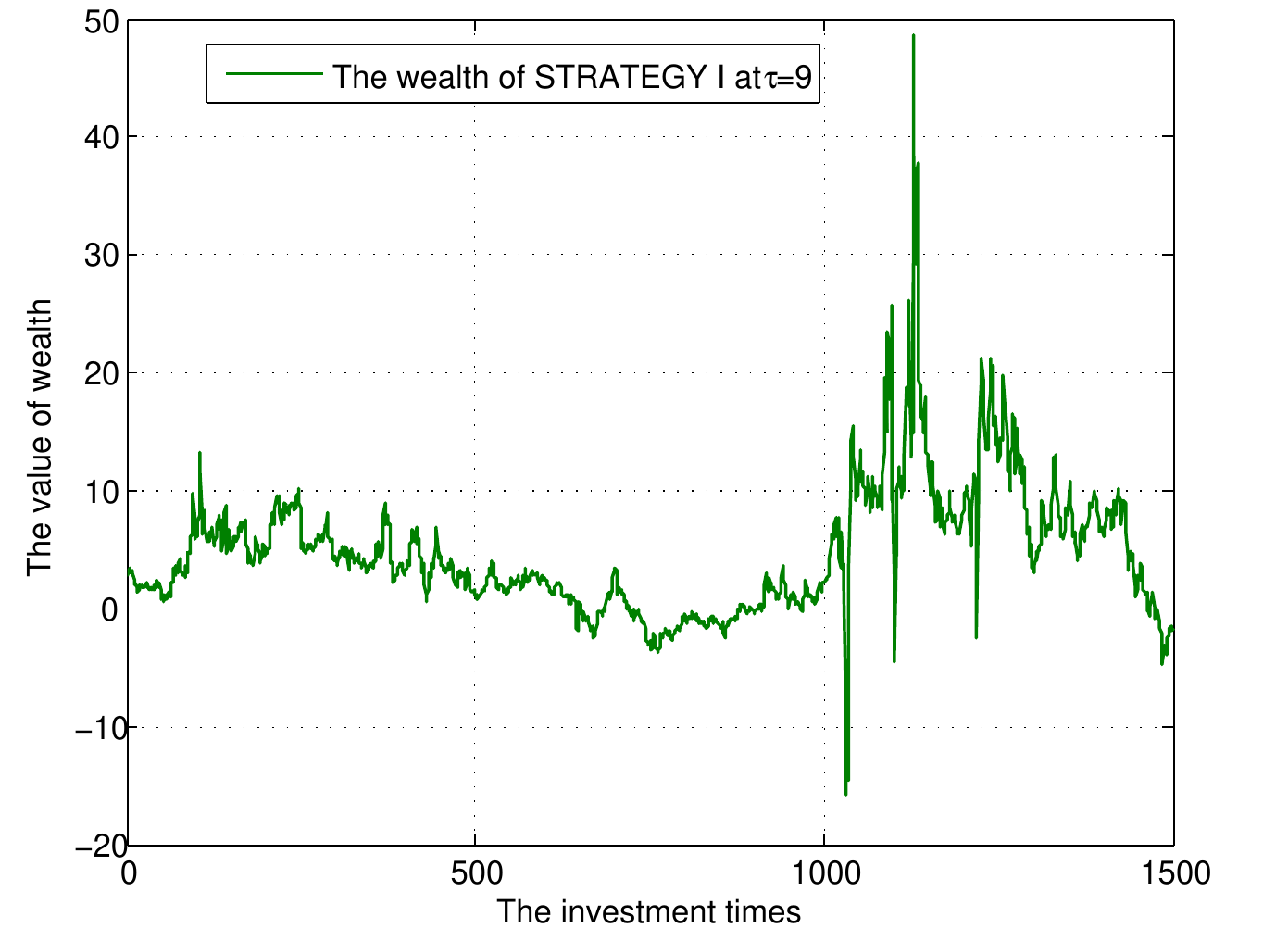}
\end{center}
\end{figure}
In Figure \ref{fig5}, we repeat the  multi-period investment portfolio of STRATEGY I  $1500$ times along with $w\leq t\leq w+1499$.
Here, we can see that the wealth of STRATEGY I is always larger than 0, and the volatility of wealth is large, relative to the initial wealth $x=1$.
 Thus we conclude the average yearly return ${\mathrm{Return}}_1(\tau,L)$ and Sharpe ratio ${\mathrm{Sharpe}}_1(\tau,L)$ of STRATEGY I from Jan. 2012 to Jan. 2018 as follows:
\begin{table}[H]
  \centering
  \caption{Average yearly return and Sharpe ratio of STRATEGY I with $L=30$, transaction fee rate $r_0=0.001$ and daily loan interest rate $\overline{r}=0.0003$}
  \label{tab-6}
  \begin{tabular}{lcccccc}
    \toprule
    Investment times    & Initial wealth    & Investment period &  Yearly return  &  Sharpe ratio  \\
    \midrule
    Jan. 2012--Jan. 2013 &$x=1$    &$\tau=9$  &  $4.1683$  & $1.4993$\\
    Jan. 2013--Jan. 2014 &$x=1$    &$\tau=9$  &  $3.2769$  & $2.0746$\\
    Jan. 2014--Jan. 2015 &$x=1$    &$\tau=9$  &  $0.0574$  & $0.0047 $\\
    Jan. 2015--Jan. 2016 &$x=1$    &$\tau=9$  &  $-1.3317$ & $-0.8853$\\
    Jan. 2016--Jan. 2017 &$x=1$    &$\tau=9$  &  $9.9944$  & $1.2571 $\\
    Jan. 2017--Jan. 2018 &$x=1$    &$\tau=9$  &  $5.4866$  & $1.1085$\\
    The average value    &$x=1$    & ---      &  $3.6086$  & $0.8432$              \\
    \bottomrule
    \hline
  \end{tabular}
\end{table}

For the given daily loan interest rate $\overline{r}=0.0003$, we show how the transaction fee rate $r_0\in\{0.001,0.002,\cdots,0.010\}$ affects the average yearly return and Sharpe ratio of  portfolio wealth of STRATEGY I:
\begin{table}[H]
  \centering
  \caption{Yearly return and Sharpe ratio of STRATEGY I with $L=30$ and investment period $\tau=9$}
  \label{tab-7}
  \begin{tabular}{lcccccc}
    \toprule
    Investment times    & Initial wealth    & Transaction fee rate  &  Yearly return  &  Sharpe ratio  \\
    \midrule
    Jan. 2012--Jan. 2018 &$x=1$    &$r_0=0.001$  &  $3.6086$  & $0.6332$\\
    Jan. 2012--Jan. 2018 &$x=1$    &$r_0=0.002$  &  $3.1921$  & $0.5785$\\
    Jan. 2012--Jan. 2018 &$x=1$    &$r_0=0.003$  &  $2.7756$  & $0.5193 $\\
    Jan. 2012--Jan. 2018 &$x=1$    &$r_0=0.004$  &  $2.3591$  & $ 0.4551$\\
    Jan. 2012--Jan. 2018 &$x=1$    &$r_0=0.005$  &  $1.9426$  & $0.3856$\\
    Jan. 2012--Jan. 2018 &$x=1$    &$r_0=0.006$  &  $1.5261$  & $0.3107$\\
    Jan. 2012--Jan. 2018 &$x=1$    &$r_0=0.007$  &  $1.1096$  & $0.2302 $\\
    Jan. 2012--Jan. 2018 &$x=1$    &$r_0=0.008$  &  $0.6931$  & $ 0.1439$\\
    Jan. 2012--Jan. 2018 &$x=1$    &$r_0=0.009$  &  $0.2766$  & $ 0.0521 $\\
    Jan. 2012--Jan. 2018 &$x=1$    &$r_0=0.010$  &  $-0.1400$ & $-0.0449$\\
    \bottomrule
    \hline
  \end{tabular}
\end{table}

From Table \ref{tab-7}, we can see that the average yearly return and Sharpe ratio of the wealth is decreasing with the value of transaction fee rate $r_0$. Note here that, we have not introduced the transaction fee rate when deriving STRATEGY I, and considering the transaction fee in the empirical analysis. It is interesting to find an optimal strategy for the discrete time multi-period mean-variance model with transaction fee rate. However, we can see that the yearly return is larger than $1.9426$ and Sharpe ratio is larger than $0.3856$ when the transaction fee rate $r_0\leq 0.005$ under the initial wealth $x=1$. Additionally, we want to highlight that STRATEGY I is the proportional investment of the initial wealth $x$, which means that the average yearly return and Sharpe ratio of per unit wealth will not change with the value of initial wealth $x$.

\section{Conclusion}\label{con}
By introducing a deterministic process $Y^{\pi}(\cdot)=\mathbb{E}[X^{\pi}(\cdot)]$ with a initial value $y$, we consider the following value function:
 \begin{equation}\label{c-cost}
V^{\mu}(t,x,y)=\inf_{\pi\in \mathcal{A}_{t}^{T-1}}\tilde{J}(t,x,y,\mu;\pi(\cdot))=\inf_{\pi\in \mathcal{A}_{t}^{T-1}} \mathbb{E}[\mu\big(X^{\pi}(T)-Y^{\pi}(T)\big)^2-X^{\pi}(T)],
\end{equation}
where $X^{\pi}(T)$ with the initial value $x$. From the cost functional (\ref{c-cost}), we can distinguish the wealth process $X^{\pi}(\cdot)$ and process $Y^{\pi}(\cdot)$. Based on these setting, we can derive the related dynamic programming principle for the value function $V^{\mu}(t,x,y)$. The main results of this study are given as follows:
\begin{itemize}
\item  Similar with the idea in \cite{Y20}, we solve a variance type cost functional that contains a nonlinear part of the mean process $\mathbb{E}[X^{\pi}(\cdot)]$ in discrete time case. This new method can help us to separate the nonlinear part of the mean process from the variance in cost functional, and then we can obtain the related dynamic optimal strategy which is time consistent.

 \item Furthermore, we develop a varying investment period discrete time multi-period mean-variance model, for which we obtain the related dynamic time-consistent optimal strategy and optimal investment period.

 \item To compare our dynamic optimal strategy with the $1/n$ equality strategy, we use the daily data of NASDAQ and Dow Jones to construct portfolio investment for our dynamic optimal strategy and the $1/n$ equality, which demonstrates that our optimal strategy is better than  $1/n$ equality strategy at least for index data NASDAQ and Dow Jones in empirical analysis.

\end{itemize}

\appendix
\section{Supplementary of the main proofs}\label{app}

\noindent\textbf{The proof of Theorem \ref{the-1}}:  Based on the method in the proof of Theorem 3.3, Chapter 4 in \cite{YZ99}, we can prove this result, also see \cite{Y20}. For reader's convenience, we show the details of the proof. For any given $0\leq t\leq s< T,\ x,y\in \mathbb{R}$, the value function is given as follows:
\begin{equation}
V^{\mu}(t,x,y)=\inf_{\pi(\cdot)\in \mathcal{A}^{T-1}_t}\tilde{J}(t,x,y,\mu;\pi(\cdot)),
\end{equation}
where
\begin{equation*}
\begin{array}{rl}
\tilde{J}(t,x,y,\mu;\pi(\cdot))=\mu \mathbb{E}[\big(X^{\pi}(T)-Y^{\pi}(T)\big)^2]-\mathbb{E}[X^{\pi}(T)].
\end{array}
\end{equation*}
Note that $Y^{\pi}(\cdot)$ is a deterministic process, for the cost functional $\tilde{J}(t,x,y,\mu;\pi(\cdot))$, we can obtain
\begin{equation}\label{dpp-1}
\tilde{J}(s,X^{\pi}(s),Y^{\pi}(s),\mu;\pi(\cdot))=\mathbb{E}\big[\mu\big(X^{\pi}(T)-Y^{\pi}(T)\big)^2
-X^{\pi}(T)\ \big| \ \mathcal{F}_s \big].
\end{equation}

We first set,
\begin{equation*}
\tilde{V}^{\mu}(t,x,y):=\inf_{\pi(\cdot)\in \mathcal{A}_t^{s-1}}\mathbb{E}[V^{\mu}(s,X^{\pi}(s),Y^{\pi}(s))].
\end{equation*}
For any given $\varepsilon>0$, there exists $\pi^{\varepsilon}(\cdot)$ such that
\begin{equation}
\begin{array}{rl}
V^{\mu}(t,x,y)+\varepsilon>& \tilde{J}(t,x,y,\mu;\pi^{\varepsilon}(\cdot))\\
=& \mathbb{E}[\mu\big(X^{\pi^{\varepsilon}}(T)-Y^{\pi^{\varepsilon}}(T)\big)^2-X^{\pi^{\varepsilon}}(T)] \\
=&\mathbb{E}\big[\mathbb{E}\big[\mu\big(X^{\pi^{\varepsilon}}(T)-Y^{\pi^{\varepsilon}}(T)\big)^2
-X^{\pi^{\varepsilon}}(T)\ \big| \ \mathcal{F}_s \big]\big]\\
=&\mathbb{E}\big[\tilde{J}(s,X^{\pi^{\varepsilon}}(s),Y^{\pi^{\varepsilon}}(s),\mu;\pi^{\varepsilon}(\cdot))\big]\\
\geq &\mathbb{E}\big[ {V}^{\mu}(s,X^{\pi^{\varepsilon}}(s),Y^{\pi^{\varepsilon}}(s)) \big]\\
\geq &\tilde{V}^{\mu}(t,x,y).\\
\end{array}
\end{equation}
Thus $V^{\mu}(t,x,y)+\varepsilon>\tilde{V}^{\mu}(t,x,y)$. In contrast, we prove $V^{\mu}(t,x,y)<\tilde{V}^{\mu}(t,x,y)+\varepsilon$ in the following. Note that, for any given
$\pi(\cdot)\in \mathcal{A}_t^{T-1}$,
\begin{equation}
\left\{\begin{array}{rl}
X^{\pi}(s)  & \!\!\!=r(s-1)X^{\pi}(s-1)  +\gamma(s-1)\pi(s-1)^{\top}+\pi(s-1)\sigma(s-1) \Delta W(s-1),  \\
\!X^{\pi}(t) & \!\!\!=x,\ \ t<s\leq T.
\end{array}\right.
\end{equation}
There exists $\delta>0$ such that
$
\left|x_1-x_2\right|+\left|y_1-y_2\right|<\delta,
$
it follows,
$$
\left|\tilde{J}(t,x_1,y_1,\mu;\pi(\cdot))-\tilde{J}(t,x_2,y_2,\mu;\pi(\cdot))\right|+
\left|V^{\mu}(t,x_1,y_1)-V^{\mu}(t,x_2,y_2) \right|<\frac{\varepsilon}{3}.
$$
Thus, we can find a strategy
\begin{eqnarray*}
\hat{\pi}(h)=\left\{\begin{array}{ll}
 \pi(h),\quad t\leq h< s,\\
\tilde{\pi}(h),\quad s\leq h< T,
\end{array}\right.
\end{eqnarray*}
where $\pi(\cdot)\in \mathcal{A}_t^{s-1}$ is a any given strategy, such that
$$
\tilde{J}(s,X^{{\pi}}(s),Y^{{\pi}}(s),\mu;\tilde{\pi}(\cdot))<
V^{\mu}(s,X^{{\pi}}(s),Y^{{\pi}}(s))+\varepsilon.
$$
Thus, for the strategy $\hat{\pi}(\cdot)$, we have
\begin{equation}\label{app-hj-2}
\begin{array}{rl}
&{V}^{\mu}(t,x,y)\\
\leq &\displaystyle
\mathbb{E}[\mu\big(X^{\hat{\pi}}(T)-Y^{\hat{\pi}}(T)\big)^2-X^{\hat{\pi}}(T)] \\
=&\displaystyle   \mathbb{E}\big[\mathbb{E}\big[\mu\big(X^{\hat{\pi}}(T)-Y^{\hat{\pi}}(T)\big)^2-X^{\hat{\pi}}(T)  \ |\ \mathcal{F}_s\big]\big]                           \\
=&\displaystyle  \mathbb{E}\big[
\tilde{J}(s,X^{{\pi}}(s),Y^{{\pi}}(s),\mu;\hat{\pi}(\cdot)) \big]
\\
< & \displaystyle  \mathbb{E}\big[V^{\mu}(s,X^{{\pi}}(s),Y^{{\pi}}(s))\big]+\varepsilon,
\end{array}
\end{equation}
for $\pi(\cdot)\in \mathcal{A}_t^{s-1}$ is a any given strategy, we have
\begin{equation}\label{app-hj-3}
V^{\mu}(t,x,y)\leq \tilde{V}^{\mu}(t,x,y)+\varepsilon,
\end{equation}
which completes the proof. $\quad \qquad \Box$

\bigskip

\appendix
\section{Proofs}
\label{sec:proof}

\noindent\textbf{The proof of Theorem \ref{the-2}}: For any given $x,y\in \mathbb{R}$, we have the following result for time $T-1$,
\begin{equation*}
\begin{array}{rl}
 V^{\mu}(T-1,x,y)=&\displaystyle \inf_{\pi(\cdot)\in \mathcal{A}^{T-1}_{T-1}} \mathbb{E}[\mu\big(X^{\pi}(T)-Y^{\pi}(T)\big)^2-X^{\pi}(T)]\\
 =&\displaystyle\inf_{\pi(\cdot)\in \mathcal{A}^{T-1}_{T-1}}\mathbb{E}\bigg[\mu\pi(T-1)[\sigma(T-1)\sigma(T-1)^{\top}]\pi(T-1)^{\top}\\
& -r(T-1)x-\gamma(T-1)\pi(T-1)^{\top}\\
 &+\mu \big[r(T-1)(x-y)+\gamma(T-1)(\pi(T-1)^{\top}-\mathbb{E}[\pi(T-1)^{\top}])\big]^2\bigg].\\
\end{array}
\end{equation*}
Thus, we can obtain the optimal strategy,
$$
\pi^*(T-1,x,y)=\frac{\gamma(T-1)}{2\mu} {[\sigma(T-1)\sigma(T-1)^{\top}]^{-1}},
$$
and the related value function,
$$
V^{\mu}(T-1,x,y)=r(T-1)\big[\mu r(T-1)(x-y)^2-x\big]-\frac{\beta(T-1)}{4\mu}.
$$
Thus, it follows that
\begin{equation*}
\begin{array}{rl}
&V^{\mu}(T-1,X^{\pi}(T-1),Y^{\pi}(T-1))\\
=&\displaystyle \mu r^2(T-1)(X^{\pi}(T-1)-Y^{\pi}(T-1))^2
-r(T-1)X^{\pi}(T-1)-\frac{\beta(T-1)}{4\mu}.\\
\end{array}
\end{equation*}

For time $T-2$, we have
\begin{equation*}
\begin{array}{rl}
 V^{\mu}(T-2,x,y)=&\displaystyle \inf_{\pi(\cdot)\in \mathcal{A}^{T-2}_{T-2}} \mathbb{E}[V^{\mu}(T-1,X^{\pi}(T-1),Y^{\pi}(T-1))]\\
 =&\displaystyle\inf_{\pi(\cdot)\in \mathcal{A}^{T-2}_{T-2}}\mathbb{E}\bigg[\mu r^2(T-1)\pi(T-2)[\sigma(T-2)\sigma(T-2)^{\top}]\pi(T-2)^{\top}\\
& -r(T-1)r(T-2)x-r(T-1)\gamma(T-2)\pi(T-2)^{\top}\\
 &+\mu r^2(T-1)\big[r(T-2)(x-y)+\gamma(T-2)(\pi(T-2)^{\top}-\mathbb{E}[\pi(T-2)^{\top}])\big]^2\bigg]\\
 &\displaystyle -\frac{\beta(T-1)}{4\mu}.
\end{array}
\end{equation*}
The optimal strategy is
$$
\pi^*(T-2,x,y)=\frac{1}{2\mu}\frac{\gamma(T-1)}{r(T-1)} {[\sigma(T-1)\sigma(T-1)^{\top}]^{-1}},
$$
and the related value function is
$$
V^{\mu}(T-2,x,y)=\mu \prod_{s=T-2}^{T-1} r^2(s)(x-y)^2-\prod_{s=T-2}^{T-1}r(s)x
-\frac{\sum_{s=T-2}^{T-1}\beta(s)}{4\mu}.
$$

By Theorem \ref{the-1}, we have
\begin{equation*}
\begin{array}{rl}
 V^{\mu}(t,x,y)=&\displaystyle \inf_{\pi(\cdot)\in \mathcal{A}^{T-1}_t} \mathbb{E}[\mu\big(X^{\pi}(T)-Y^{\pi}(T)\big)^2-X^{\pi}(T)]\\
 =&\displaystyle \inf_{\pi(\cdot)\in \mathcal{A}^{T-2}_t}\inf_{\pi(\cdot)\in \mathcal{A}^{T-1}_{T-1}}\mathbb{E}[\mu\big(X^{\pi}(T)-Y^{\pi}(T)\big)^2-X^{\pi}(T)]\\
 =&\displaystyle \inf_{\pi(\cdot)\in \mathcal{A}^{T-2}_t}\mathbb{E}[V^{\mu}(T-1,X^{\pi}(T-1),Y^{\pi}(T-1))]\\
 =&\displaystyle \inf_{\pi(\cdot)\in \mathcal{A}^{T-3}_t}\inf_{\pi(\cdot)\in \mathcal{A}^{T-2}_{T-2}}\mathbb{E}[V^{\mu}(T-1,X^{\pi}(T-1),Y^{\pi}(T-1))]\\
 &\cdots \cdots\\
 =& \displaystyle \inf_{\pi(\cdot)\in \mathcal{A}^{t}_t}
 \mathbb{E}[V^{\mu}(t+1,X^{\pi}(t+1),Y^{\pi}(t+1))].\\
\end{array}
\end{equation*}
Therefore, we can employ the above method to obtain the optimal strategy for $0\leq t<T$,
$$
\pi^*(t,x,y)=\frac{\gamma(t)}{2\mu\prod_{s=t}^{T-1} r(s)} {[\sigma(t)\sigma(t)^{\top}]^{-1}},
$$
and the related value function
$$
V^{\mu}(t,x,y)=\mu(x-y)^2 \bigg(\prod_{s=t}^{T-1} r(s)\bigg)^2-x\prod_{s=t}^{T-1}r(s)
-\frac{\sum_{s=t}^{T-1}\beta(s)}{4\mu}.
$$
This completes this proof. $\quad \qquad \Box$

\bigskip

\noindent {\textbf{Proof of Proposition \ref{pro-1}.}} For a given mean level $L>\prod_{h=t}^{T-1}r(h)$ in constrained condition (\ref{mean-1}). The optimal strategy $\pi^*(\cdot)$ and $\pi^*_0(\cdot)$ satisfy
$$
\mathbb{E}[{X}^{{\pi}^{*}}(T)]=\mathbb{E}[{X}^{{\pi}^{*}_0}(T)]=L.
$$
By formulations (\ref{eff-1}) and (\ref{pre-eff}), we have
$$
\displaystyle \mathrm{Var}[X^{{\pi}^{*}}(T)]=
\frac{\bigg(
\mathbb{E}[X^{{\pi}^{*}}(T)]-x\prod_{h=t}^{T-1}r(h)\bigg)^2}{\sum_{h=t}^{T-1}\beta(h)}
$$
and
$$
\displaystyle  \mathrm{Var}[X^{{\pi}_0^{*}}(T)]=
\frac{\bigg{(}\mathbb{E}[X^{{\pi}_0^{*}}(T)]
-x\prod_{h=t}^{T-1}r(h)\bigg{)}^2}{\prod_{h=t}^{T-1}[\beta(h)+1]-1}.
$$
By Assumption $\mathrm{\textbf{H}}_2$, we have $\beta(s)>0,\ t\leq s< T$, and
$$
\sum_{h=t}^{T-1}\beta(h)<\prod_{h=t}^{T-1}[\beta(h)+1]-1.
$$
Therefore, one obtains,
$$
\mathrm{Var}[X^{{\pi}^{*}}(T)]> \mathrm{Var}[X^{{\pi}_0^{*}}(T)].
$$

For a given risk aversion parameter $\mu>0$, we have
$$
\mathbb{E}[X^{{\pi}^{*}}(T)]=\displaystyle  x\prod_{h=t}^{T-1}r(h)+\sum_{h=t}^{T-1}\frac{\beta(h)}{2\mu},
$$
and
$$
\displaystyle \mathbb{E}[{X}^{{\pi}^{*}_0}(T)]=
x \prod_{h=t}^{T-1}r(h)+\frac{1}{2\mu}(\prod_{h=t}^{T-1}[\beta(h)+1]-1).
$$
From $\beta(s)>0,\ t\leq s< T$, it follows
$$
\frac{1}{2\mu}\sum_{h=t}^{T-1}\beta(h)<\frac{1}{2\mu}\bigg(\prod_{h=t}^{T-1}[\beta(h)+1]-1\bigg),
$$
which implies that
$$
x \prod_{h=t}^{T-1}r(h)
<\mathbb{E}[X^{{\pi}^{*}}(T)]<\mathbb{E}[{X}^{{\pi}^{*}_0}(T)].
$$
Again, by formulations (\ref{eff-1}) and (\ref{pre-eff}), we have
$$
\mathrm{Var}[X^{{\pi}^{*}}(T)]=\frac{\sum_{h=t}^{T-1}\beta(h)}{4\mu^2}<
\frac{\prod_{h=t}^{T-1}[\beta(h)+1]-1}{4\mu^2}=\mathrm{Var}[X^{{\pi}_0^{*}}(T)].
$$
Therefore,
\begin{equation}\label{app-pro-1-var}
\mathrm{Var}[X^{{\pi}^{*}}(T)]<\mathrm{Var}[X^{{\pi}_0^{*}}(T)],\quad \mathbb{E}[{X}^{{\pi}^{*}}(T)]<\mathbb{E}[{X}^{{\pi}^{*}_0}(T)].
\end{equation}
This completes the proof. $\quad \qquad \Box$

\bibliography{var}
\end{document}